\documentclass[12pt]{elsarticle}
\usepackage{amsmath}
\usepackage[usenames, dvipsnames]{xcolor}
\usepackage[utf8x]{inputenc}
\usepackage[T1]{fontenc}
\usepackage{tikz}
\usepackage{pgfplots, pgfplotstable}
\usetikzlibrary{calc,arrows,fadings,decorations.pathreplacing,decorations.markings,patterns,shapes.geometric}
\usepackage{fp}
\usepackage{yhmath}
\usepackage{verbatim}
\usepackage{amssymb}
\usepackage{caption}
\usepackage{subcaption}
\usetikzlibrary{calc,fit,intersections,shapes}
\usetikzlibrary{calc}
\usepgfplotslibrary{polar}
\pgfplotsset{compat=1.6}
\pgfplotsset{every axis plot}
\journal{ }
\usepackage{natbib}
\usepackage{amsmath}
\usepackage[normalem]{ulem}
\usepackage{multirow}

\begin{document}

\title{Design optimisation of piezoelectric energy harvesters for bridge infrastructure}

\author[label1]{P. Peralta-Braz}
\author[label1]{M. M. Alamdari}
\author[label2]{R. O. Ruiz}
\author[label1]{E. Atroshchenko\footnote{Corresponding author, e.atroshchenko@unsw.edu.au, eatroshch@gmail.com}}
\author[label3]{M. Hassan}

\address[label1]{School of Civil and Environmental Engineering, University of New South Wales, Sydney, Australia}
\address[label2]{Department of Civil Engineering, Universidad de Chile, Av. Blanco Encalada 2002, Santiago,Chile}
\address[label3]{School of Computer Science and Engineering, University of New South Wales, Sydney, Australia}

\journal{Mechanical Systems and Signal Processing}
\begin{abstract}
Vibrational energy harvested from the bridge excitation due to the traffic flow or the wind load can be supplied to sensors in a Structural Health Monitoring (SHM) system and help prolong its service life, reduce chemical battery waste and enable its use in remote locations. A common approach to designing a Piezoelectric Energy Harvester (PEH) consists in tuning its fundamental frequency to some target value (e.g., the fundamental frequency of a bridge). However, such approach does not answer the question if the chosen target frequency is optimal, and does not take into account the possibility to have multiple design configurations with the same fundamental frequency. In this work, we approach the problem of a PEH design optimisation in a rigorous way, using a PEH model based on the Kirchhoff-Love plate theory and Isogeometric Analysis (IGA), coupled with the Modal Order Reduction (MOR) approach and Runge-Kutta time integration method. The model is further equipped with the Particle Swarm Optimisation (PSO) algorithm that allows finding geometry which maximises energy output from a given base acceleration signal. A comprehensive study is conducted to infer the impact of a PEH geometry on its energy harvesting performance in a real-world setting by considering field health monitoring data of a large-scale cable-stayed bridge in NSW, Australia. A shape optimisation framework is developed based on acceleration events, i.e., where the level of response exceeds a certain threshold. Designs obtained in the event-based optimisation are then clustered to propose several best candidates for continuous energy generation. This work represents the first study on PEH design optimisation for real operational conditions. 
  
\end{abstract}

\maketitle

\section{Introduction}
\label{S:1}
Bridges are crucial elements in transportation systems due to their cost, the connectivity they add to the network, and the severe consequences of their collapse. For this reason, Structural Health Monitoring (SHM) systems are widely used to ensure bridges' safe operation and, in particular, to prevent overloading, which is the principal cause of deterioration and surface damage. SHM systems require implementing an array of sensors, which are commonly supplied through connecting cables. Nonetheless, cabling can be an inflexible, expensive, and unsafe procedure. Alternatively, wireless sensors powered by chemical batteries are used. However, battery replacement can be particularly difficult, expensive, or unfeasible in hard to reach areas, limiting the service life of a SHM system to the service life of a battery and producing substantial amount of chemical waste. Recently, vibration energy harvesting has become a promising alternative to provide energy supply for sensors on bridges, taking advantage of kinetic energy induced by the wind load and traffic flow. In particular, Piezoelectric Energy Harvesters (PEHs) \cite{covaci2020piezoelectric, sezer2021comprehensive} are a widespread and popular option among vibrational energy harvesting systems due to their high electrical power density. PEHs are able to convert vibrational energy into electrical power using the direct piezoelectric effect \cite{wu2021review}. The most common configuration of PEHs consists of a cantilevered plate composed of a substructure layer and one or two piezoelectric layers. 

Studies in the literature agree on the potential feasibility of using PEH to supply electrical power to electronics commonly used for structural health monitoring of bridges. In the work of Karimi el al. \cite{karimi2016experimental}, the reported root mean square voltage is between 0.4-8.0 mV. The study indicates that the voltage is almost proportional to the speed increment of the vehicle. Peigney and Siegert \cite{peigney2013piezoelectric} reported that for the peak intensity of traffic, the power reaches 0.03 mW with a voltage between 1.8 and 3.6 V, indicating that it is enough to supply wireless health monitoring nodes with a low duty cycle. In the work of Zhang et al. \cite{zhang2018experimental}, the higher device's efficiency was achieved when it was installed in the middle of the bridge. The PEHs harvested around 579 $\mu$J energy each time a vehicle passed over the bridge. In the research by Song \cite{song2019finite}, the root mean square value of the output voltage varies in the interval 0.5–205 V among all the scenarios of bridge vibrations. The energy varies in the interval 8.0-462 $\mu$J. The authors indicate, that in the most favourable case, the energy scavenged by one PEH is enough to power a wireless intelligent sensor and an actuator network node. Romera et al. \cite{romero2021energy} report that the energy generated from a time window of three and a half hours and twenty train passages is 3.6 mJ. 

Results presented in \cite{peigney2013piezoelectric} and \cite{zhang2018experimental} are based on the experimental studies. Although the experimental approach provides a reliable estimation of PEH's performance under real operating conditions, it is limited to testing only a small number of devices since each device has to be designed and manufactured. Most common design approach consists in synchronising the resonant frequency of the device with one of the natural frequencies of the bridge. In reference \cite{peigney2013piezoelectric}, the prototype's resonance frequency was designed to target the resonance vibration of a piping structure attached to the bridge. In reference \cite{zhang2018experimental}, the two PEHs were designed with different fundamental frequencies, one with the bridge's natural frequency and the other with the frequency of a vehicle-bridge coupling vibration. While this can lead to satisfactory results, they are not necessarily optimal, and they do not reflect the full potential of a PEH to harvest energy from the bridge vibration. To elaborate on this issue, Peigney and Siegert \cite{peigney2013piezoelectric} proposed a formulation that relates the produced power to the traffic intensity. The authors state that the studied piezoelectric device is not optimal, opening the discussion of designing a PEH that maximises energy production. One more study worth mentioning in this context is the experiment done in \cite{cahill2018vibration, cahill2018data}, where the two sets of PEHs were designed with resonance frequencies to target the first two modes of a railway bridge. The authors studied the magnitude of the generated voltage, and also explored the use of the voltage signal to perform modal identification and train passage detection. Field tests were performed on the harvesters installed on the bridge while a shaker mounted below the bridge deck induced a swept sinusoidal force at different amplitudes and frequency ranges. Authors show that devices of different geometries produce different amount of energy and show different performance in monitoring tasks (modal identification and train passage detection). Hence, these results confirm the importance of a PEH design on both, energy generation and sensing performance. 

Studies in \cite{karimi2016experimental, song2019finite, romero2021energy} have addressed the use of PEHs on bridges using a theoretical approach, based on mathematical and numerical models for the vehicle-bridge interaction coupled with the PEH. Despite the fact that a numerical model provides greater versatility in designing and selecting a device in comparison with experimental studies, to the best of our knowledge, no studies have presented a framework for optimisation of PEH design for bridges, even though previous works have shown that the shape of the device has a high impact on its performance \cite{savarimuthu2018design, hosseini2016shape, peralta2020parametric, hurtado2022shape}.

A common approach to modelling PEHs in the literature is based on the Euler-Bernoulli beam theory, even though PEHs are designed and manufactured as thin structures more similar to plates than beams, making plate theories more suitable. The first PEH model based on the Kirchhoff-Love plate and Hamilton's generalised principle for electrostatic bodies was presented in \cite{junior2009electromechanical}. The model was solved numerically using the finite element method, and the obtained results showed an excellent agreement with the experimental data. Later, Peralta et al. \cite{peralta2020parametric} presented an Iso-Geometric Analysis (IGA) framework for a PEH based on the Kirchhoff-Love plate, additionally equipped with a shape optimisation algorithm to maximise the amount of produced voltage. The two main advantages of the IGA-PEH model are its capacity to parameterise the shape of the device by B-Splines, allowing complex configurations with many design parameters, and its high accuracy due to the high order and high continuity of the B-Spline basis functions. In addition, in terms of computational time, the IGA-based framework is faster than FEM analysis with the same accuracy, which is an essential advantage in the optimisation process that requires solving the problem for a large number of iterations. Later, in \cite{hurtado2022shape}, the model was extended to devices of variable thickness. Thus, the IGA PEH model is chosen for the present study. 

To summarise the above discussion, we can say that the studies on the use of PEHs for bridges suggest that storing the produced energy and using it in continuous or intermittent measurement operations is feasible. However, the PEH design process has not been addressed in-depth in the literature. In general, design procedure is limited to matching the resonant frequency of the device with some target value (e.g., one of the natural frequencies of the bridge or the frequency of a vehicle-bridge coupling vibration), however, to the best of our knowledge, no rigorous and systematic studies exist to justify this approach. Therefore, proposing rigorous and versatile design optimisation framework to maximise the energy harvesting efficiency on bridges can contribute to the widespread use of PEHs.

The contributions of this work are summarised as follows:

\begin{itemize}
    \item The IGA-PEH model, developed in \cite{peralta2020parametric} for a single harmonic acceleration input, is extended for an arbitrary base acceleration using the Modal Order Reduction and Runge-Kutta time integration method. The model is also coupled with the Particle Swarm Optimisation (PSO) algorithm to find geometric configurations with the maximum energy output. The current study is limited to rectangular geometries with constant thickness, but the extension to more complex designs parameterised by B-Splines is straightforward. To the best of our knowledge, this is the most accurate and versatile PEH model available in the literature.
    \item A rigorous and versatile design optimisation framework is proposed. The approach consists of obtaining optimal designs for a representative number of events, then clustering optimal geometries to choose several best candidates, whose performance is subsequently assessed on a continuous 24 hours acceleration input. 
    \item Results of event-based optimisation reveal that the optimal PEH geometries depend on the spectral characteristics of an event, i.e. fundamental frequency of an optimal PEH device is tuned with the predominant frequency of the event spectrum, and in some cases, out of several devices with the same fundamental frequency, optimal geometry is determined by the second natural frequency.       
    \item Design optimisation is performed based on the field SHM data from a cable-stayed bridge in New South Wales (NSW), Australia. This work presents the first study on PEH design optimisation for a large scale bridge under real-world operating conditions. 
\end{itemize}

The remainder of the paper is organised as follows. The bridge used as a case study in this work is presented in Section 2. The theoretical framework of the PEH model for an arbitrary base acceleration using the Modal Order Reduction and time integration based on Runge-Kutta method is presented in Section 3. Next, an optimisation framework of PEH for a bridge based on time series excitation is presented in Section 4. In Section 5, preliminary results are derived to support the proposed optimisation framework. Finally, a study case is employed in Section 6 to show the framework's potential. Conclusions are presented in Section 7.

\section{Cable-Stayed Bridge} 
\label{S:B}
\begin{figure}[h!]
	\centering
	\includegraphics[width=0.50\textwidth]{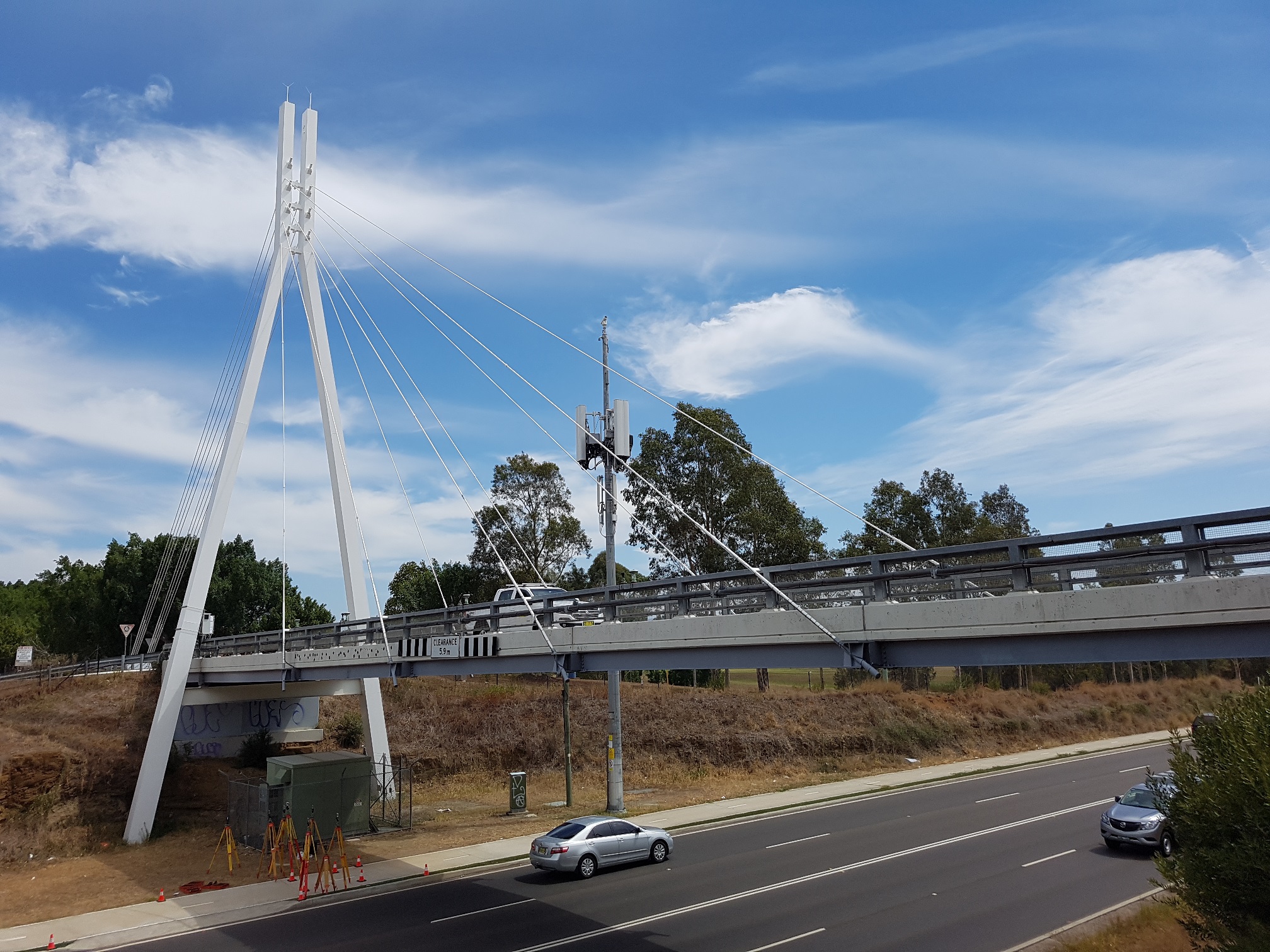}
    \caption{Illustration of the cable-stayed bridge located in the state of NSW, Australia.}
	\label{fig:fig1}
\end{figure}

\begin{figure}[h!]
	\centering\includegraphics[width=0.9\linewidth]{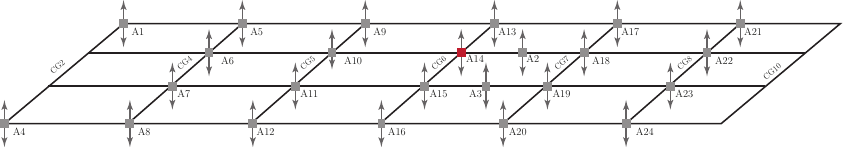}
	\caption{Illustration of the sensor layout under the deck. Acceleration response at sensor A14 was chosen for the study.}
	 \label{fig:sensor}
\end{figure}

The cable-stayed bridge shown in Figure \ref{fig:fig1} and analysed in reference \cite{alamdari2019multi} is chosen as a case study in this work. The bridge has one traffic lane and one pedestrian lane with a maximum load capacity of 30 tons over the Great Western Highway in the state of New South Wales (NSW), Australia. The structural system consists of a composite steel-concrete deck spanning over 46.2m and an A-shaped steel pylon with a height of 33m. The width and the depth of the concrete deck are uniform at 6.30m and 0.16m, respectively. The deck is supported by four longitudinal girders which are internally attached by a set of seven cross girders (CGs), see Figure \ref{fig:sensor}. Sixteen semi-fan arranged pre-tensioned stayed-cables with a diameter of 38mm are anchored on a steel tower and cross girders. Although the bridge has only one traffic lane, the vehicles can travel in both directions because the traffic flow is not high. Also, the first natural frequencies are low, i.e., 2 and 3.5 Hz for the first two modes, respectively \cite{alamdari2019damage}.\\

\begin{figure}[h!]
\centering\includegraphics[width=0.8\linewidth]{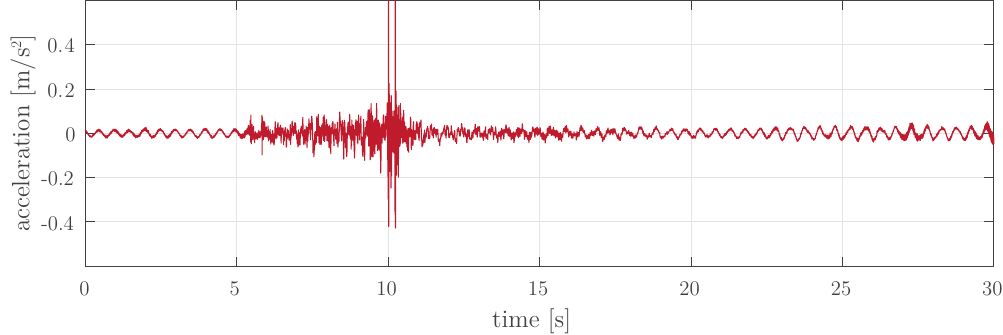}
\caption{An example of {\it an event}. This event corresponds to an acceleration time history of the cable-stayed bridge used in Sections \ref{S:Y} and \ref{S:5} as a study case. This event was labelled as event 2.}\label{event02a}
\end{figure}

In the cable-stayed bridge, an array of twenty-four uni-axial accelerometer sensors are permanently placed under the deck at the intersection of the girders and floor beams to measure the vertical acceleration of the bridge, see Figure \ref{fig:sensor}. Based on previous studies \cite{zhang2018experimental,romero2021energy}, sensor A14 is chosen for this study due to its location at the center of the bridge. In the context of this work, the acceleration time histories recorded in the bridge that exceeds a threshold will be called {\it an event}. The acceleration threshold was set at 0.15 m/s$^2$. The event time window is fixed to 30s and the location of the peak acceleration is forced to be at 10s. Note that all events will have the same time window and same location for the peak acceleration. 1000 events were extracted in a period of seven days and were labelled in order of appearance. An example of a single event (event 2) is presented in Figure \ref{event02a}. 

\section{PEH Modeling}   
\label{S:2}

\subsection{IGA Model for PEHs based on the Kirchhoff-Love Plate Theory}
\label{S:2.1}

\begin{figure}[h]
\centering\includegraphics[width=0.90\linewidth]{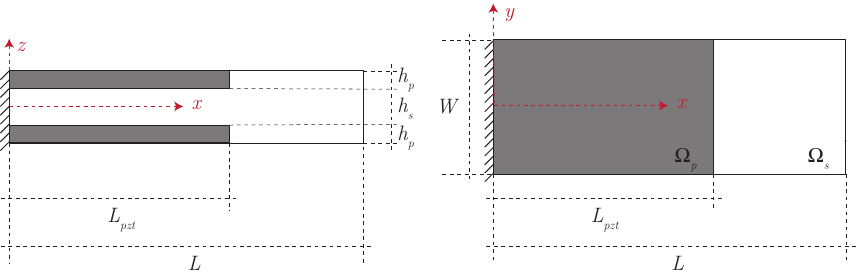}
\caption{Architecture of the piezoelectric energy harvester, corresponding to a cantilever plate, consisting of two piezoelectric layers bonded to a main substructure.}\label{PEH-scheme}
\end{figure}

The PEH modelled in this section corresponds to a bimorph-type device which is a well-known and ample studied configuration in the literature. The main architecture of the PEH corresponds to a cantilevered plate composed of three layers. The middle layer, also called substructural layer, is made of a non-piezoelectric material that serves as structural support for the plate. This layer is considered rectangular with length $L$, width $W$ and thickness $h_s$. Here, two layers made with piezoelectric ceramics are bonded to the substructural layer, as shown in Figure \ref{PEH-scheme}, and connected in series to an external electrical resistance $R_l$. The piezoelectric layers' thickness is denoted by $h_p$. Note that piezoelectric layers cover the substructure only partially (the corresponding length is denoted by $L_{pzt}$). The domains occupied by the piezoelectric layers $\Omega_{p}$ and the substructure $\Omega_{s}$ are defined by
\begin{equation}
\Omega_p = \left((0, L_{pzt})\times\left(-\frac{W}{2}, \frac{W}{2}\right)\times\left(\frac{h_s}{2}, \frac{h_s}{2} + h_p\right)\right) \cup \left((0, L_{pzt})\times\left(-\frac{W}{2}, \frac{W}{2}\right)\times\left(-\frac{h_s}{2}-h_p, -\frac{h_s}{2}\right)\right)
\end{equation}
\begin{equation}
\Omega_s = \left((0, L_{pzt})\times\left(-\frac{W}{2}, \frac{W}{2}\right)\times\left(-\frac{h_s}{2}, \frac{h_s}{2}\right)\right) \cup \left((L_{pzt}, L)\times\left(-\frac{W}{2}, \frac{W}{2}\right)\times\left(-\frac{h_s}{2}-h_p, \frac{h_s}{2}+h_p\right)\right)
\end{equation}
while the material domain of the device is given by
\begin{equation}
    \Omega = \Omega_s \cup \Omega_p
\end{equation}
Note that the most common architectures studied in the literature consider the piezoelectric layers covering the substructural layer entirely, i.e. $L_{pzt} = L$, see for instance \cite{peralta2020parametric}. Additionally, it is also common to incorporate a tip mass, $M_{tip}$, facilitating the tuning of the device at lower frequencies and increasing the vibration amplitude, hence, the energy harvested. However, in this study, we are more interested in the effect of distributed mass (due to varying geometrical parameters of the PEH plate), rather than the effect of concentrated mass (the tip mass), hence we assumed $M_{tip} = 0$. On the other hand, the motivation to work with substructural layer partially covered is to add another design variable that can potentially lead to a better performance. This is explored later in Section \ref{S:Y}. 

The mathematical model for the PEH is deduced from the generalised Hamilton's principle for electro-elastic bodies and the Kirchhoff-Love plate theory. Details of this model are given in \cite{peralta2020parametric}, hence, only the main formulation is presented here. The constitutive equation for the substructure is given by
\begin{equation}
    \textbf{T} = \textbf{c}_s \textbf{S},
\end{equation}

\noindent
where \textbf{T}, \textbf{S}, and \textbf{c}$_s$ correspond to the mechanical stress tensor, the mechanical displacement vector and the elastic stiffness matrix, respectively. On the other hand, the constitutive equation for the piezoelectric material is given by

\begin{equation}
    \begin{Bmatrix}
\textbf{T}\\
\textbf{D}
\end{Bmatrix} =
\begin{bmatrix}
\textbf{c}_p^E & -\textbf{e}^T \\
\textbf{e} & \boldsymbol{\varepsilon}^S 
\end{bmatrix}
\begin{Bmatrix}
\textbf{S}\\
\textbf{E}
\end{Bmatrix}
\end{equation}
where \textbf{D} is the electrical displacement vector, \textbf{E} is the electrical field vector, \textbf{c}$_p^E$ is the elastic stiffness matrix at constant electric field, \textbf{e} is the piezoelectric constant matrix, and $\boldsymbol{\varepsilon}^S$ is the permittivity components matrix at constant strain. 

The shape of the PEH is described by B-Splines $N_I(\boldsymbol{\xi})$ defined in the parametric space  $\boldsymbol{\xi}=(\xi,\eta)\in[0,1]\times[0,1]$, and control points $ \tilde{\textbf{x}}_I$. The deflection $w(\textbf{x}, t) = w(\boldsymbol{\xi}, t)$ is approximated, according to provisions of the isogeometric analysis, using the same basis functions $N_I(\boldsymbol{\xi})$ as follows 
\begin{equation}
    \label{eq:x}
    \textbf{x}(\boldsymbol{\xi})=\sum^{k}_{I=1}N_I(\boldsymbol{\xi})\tilde{\textbf{x}}_I;
    \  
    w(\boldsymbol{\xi}, t)=\sum_{I=1}^{k}N_I(\boldsymbol{\xi})w_I(t),
\end{equation}
where $\textbf{x}=(x,y)$ is an arbitrary material point in the neutral plane ($z = 0$), $I$ is a bivariate index, and $w_I(t)$ is the deflection projected at control point $I$ at time $t$. Substituting equation (\ref{eq:x}) and equation (\ref{eq:wh}) into the weak form of the problem, the following coupled system of algebraic equations is obtained:
\begin{equation}
\label{SistEq-1}
    \textbf{M}\ddot{\textbf{w}}+\textbf{C}\dot{\textbf{w}}+ \textbf{K}\textbf{w} - \boldsymbol{\Theta}v_p = -\textbf{F}a_b
\end{equation}
\begin{equation}
\label{SistEq-2}
    C_p\dot{v}_p + \frac{v_p}{R_l} + \boldsymbol{\Theta}^T\dot{\textbf{w}}=0 
\end{equation}

Equation (\ref{SistEq-1}) corresponds to the mechanical equation with electrical coupling while equation (\ref{SistEq-2}) corresponds to the electrical circuit equation with mechanical coupling, where $\mathbf{w}\in\mathbb{R}^{N\times 1}$ is a vector containing the deflections $w_I(t)$, $v_p=v_p(t)$ is the output voltage, $a_b=a_b(t)$ is the acceleration imposed at the base of the PEH, $\mathbf{M}\in\mathbb{R}^{N\times N}$ is the mass matrix, $\mathbf{K}\in\mathbb{R}^{N\times N}$ is the stiffness matrix, $\mathbf{C}=\alpha\mathbf{M} + \beta\mathbf{K}\in\mathbb{R}^{N\times N}$ is the mechanical damping matrix (where $\alpha$ and $\beta$ are the proportional damping coefficients), $\mathbf{F}\in\mathbb{R}^{N\times1}$ is the vector of mechanical forces, $\boldsymbol{\Theta}\in\mathbb{R}^{N\times1}$ is the electromechanical coupling vector, $C_p$ is the capacitance, and $R_l$ is the external electric resistance. The matrices $\mathbf{M}$, $\mathbf{K}$, $\mathbf{F}$ and $\boldsymbol{\Theta}$ are assembled as
\begin{equation}
    M_{IJ} = \int_{\Omega_s}\rho_s(N_IN_J+z^2N_{I,x}N_{J,x}+z^2N_{I,y}N_{J,y})d\Omega_s + 2\int_{\Omega_p}\rho_p(N_IN_J+z^2N_{I,x}N_{J,x}+z^2N_{I,y}N_{J,y})d\Omega_p
\end{equation}

\begin{equation}
    K_{IJ} = \int_{\Omega_s}z^2 \textbf{B}_I^T\textbf{c}_s\textbf{B}_J d\Omega_s + 2\int_{\Omega_p}z^2 \textbf{B}_I^T\textbf{c}_p^E\textbf{B}_J d\Omega_s 
\end{equation}
\begin{equation}
    \Theta_I = \int_{\Omega_p}z\textbf{B}_I^T\textbf{e}^T\textbf{Z} d\Omega_p
\end{equation}
\begin{equation}
\begin{split}
    {F}_I = \int_{\Omega_s}\rho_sN_I d\Omega_s &+ \int_{\Omega_p}\rho_pN_I d\Omega_p\\
             + \sum\limits_{J = 1}^{k}\int_{\Omega_s}\rho_sz^2(N_{I,x} N_{J,x} + N_{I,y} N_{J,y}) d\Omega_s &+ \sum\limits_{J = 1}^{k}\int_{\Omega_p}\rho_pz^2(N_{I,x} N_{J,x} + N_{I,y} N_{J,y}) d\Omega_p
\end{split}
\end{equation}
where $\rho_s$ is the substructure density, $\rho_p$ is the piezoelectric material density, while $\textbf{B}_I$ and $\textbf{Z}$ are defined as
\begin{equation}
    \textbf{B}_I = \begin{Bmatrix}
-N_{I,xx} & -N_{I,yy} & -2N_{I,xy}\\
\end{Bmatrix}^T;\ 
    \textbf{Z} = 
    \begin{Bmatrix}
0 & 0 & \frac{1}{h_p}\\
\end{Bmatrix}^T
\end{equation}

From equations (\ref{SistEq-1}) and (\ref{SistEq-2}), also considering that excitation and response are harmonic signals of the form $a_b(t) = A_b e^{i\omega t}$ and $v(t) = V_o e^{i\omega t}$ (where $i=\sqrt{-1}$), it is possible to define the Frequency Response Function (FRF) that relates the amplitude of the output voltage $V_o$ and the excitation acceleration $A_b$ for a specific frequency $\omega$. The voltage and power FRF are defined as follows
\begin{equation}
    \label{Hv_equation}
 H_v(\omega) = \frac{V_o(\omega)}{A_b} = i\omega\left(\frac{1}{R_l}+i\omega C_p\right)^{-1} \boldsymbol{\Theta}^T\left(-\omega^2\textbf{M}+i\omega \textbf{C} + \textbf{K} + i\omega \left(\frac{1}{R_l}+i\omega C_p\right)^{-1} \boldsymbol{\Theta} \boldsymbol{\Theta}^T\right)^{-1}\textbf{F}
\end{equation}

\begin{equation}
    \label{HP_equation}
 H_p(\omega) = \frac{H_v(\omega)^2}{R_l}
\end{equation}

The FRFs, given by equations (\ref{Hv_equation}) and (\ref{HP_equation}), characterise the dynamic response of a PEH subjected to a harmonic excitation (acceleration at the base of the PEH). However, when the PEH is intended to be used for real applications, the excitation of the PEH could potentially differ from a harmonic signal. To circumvent this limitation, it is necessary to extend the modelling process to enable the study of any arbitrary excitation, i.e., performing the time integration of equations (\ref{SistEq-1}) - (\ref{SistEq-2}). Time integration is a computationally intensive process, then, if an optimisation requires to perform a time integration to evaluate the objective function, the problem could become computationally intractable (in particular for long time series). In order to reduce this computational burden, the IGA model is coupled with a Modal Order Reduction, which is discussed next.

\subsection{Modal Order Reduction for Time Integration}
\label{S:2.2}

Here, a Modal Order Reduction (MOR) \cite{besselink2013comparison} is applied to the system described by equations (\ref{SistEq-1})- (\ref{SistEq-2}). This method allows reducing the number of degrees of freedom of the system, considering only a limited number of modes to represent the dynamic behaviour in a specific frequency range. These modes can be found from equation (\ref{SistEq-1}), considering that the device is unforced, undamped, and unpolarised, i.e.

\begin{equation}
\label{red-eq}
    \textbf{M}\ddot{\textbf{w}} + \textbf{K}\textbf{w} = \textbf{0}
\end{equation}
Then, a harmonic representation of the deflection leads to the generalised eigenvalue problem

\begin{equation}
\label{eigenvalues-problem}
    (\textbf{K} - \omega_i^2 \textbf{M})\boldsymbol{\phi}_i = \textbf{0}
\end{equation}

Here, $\boldsymbol{\phi}_i\in\mathbb{R}^{N\times1}$ represents the mode shape vectors and $\omega_i$ are the undamped natural frequencies. The deflection solution \textbf{w} can be approximated through a truncated expansion of the first $K$ mode shape vectors, associated with the first $K$ natural frequencies as

\begin{equation}
\label{deflection-approximation}
    \textbf{w}\approx\textbf{w}_o = \boldsymbol{\Phi}_o \boldsymbol{\eta}
\end{equation}
where $\textbf{w}_o\in\mathbb{R}^{N \times 1}$ is the approximate deflection solution, $\boldsymbol{\Phi}_o\in\mathbb{R}^{N \times K}$ is the matrix which contains the first $K$ mode shape vectors $\boldsymbol{\phi}_i$ and $\boldsymbol{\eta}\in\mathbb{R}^{K\times1}$ denotes the modal coordinates. In this way, it is possible to define the reduced mass matrix $\textbf{m}_o$:
\begin{equation}
\label{reduced-mass}
    \textbf{m}_o = \boldsymbol{\Phi}_o^T \textbf{M} \boldsymbol{\Phi}_o
\end{equation}

Next, equations (\ref{SistEq-1}) and (\ref{SistEq-2}) can be represented in modal coordinates and normalised by the reduced mass matrix resulting in a reduced-order system
\begin{equation}
\label{SistEq-1R}
    \ddot{\boldsymbol{\eta}}+\textbf{c}_o\dot{\boldsymbol{\eta}}+ \textbf{k}_o\boldsymbol{\eta} - \boldsymbol{\theta}_ov_p = \textbf{f}_o a_b
\end{equation}

\begin{equation}
\label{SistEq-2R}
    C_p\dot{v}_p + \frac{v_p}{R_l} + \boldsymbol{\Theta}^T\boldsymbol{\Phi}_o \dot{\boldsymbol{\eta}}=0 
\end{equation}
where

\begin{equation}
    \begin{split}
       \textbf{c}_o & = \textbf{m}_o^{-1} \boldsymbol{\Phi}_o^T \textbf{C} \boldsymbol{\Phi}_o \\
       \textbf{k}_o & = \textbf{m}_o^{-1} \boldsymbol{\Phi}_o^T \textbf{K} \boldsymbol{\Phi}_o \\
       \boldsymbol{\theta}_o & = \textbf{m}_o^{-1} \boldsymbol{\Phi}_o^T \boldsymbol{\theta} \\
       \textbf{f}_o & = \textbf{m}_o^{-1} \boldsymbol{\Phi}_o^T \textbf{F}\\
    \end{split}\label{mass_stiff_MOR}
\end{equation}


Note that the new stiffness and damping matrices, $\textbf{k}_o$ and $\textbf{c}_o$ (equation (\ref{mass_stiff_MOR})), are now defined in terms of the undamped natural frequencies $\omega_i$ and the damping ratios $\zeta_i$ associated to the modes of interest $\{i=1,...,K\}$. 



The voltage generated by the piezoelectric devices is obtained from the solution of the system of differential equations given by equations (21) and (22). The system of differential equations can be written in its state-space representation as
\begin{equation}
    \dot{\textbf{Z}} = \textbf{A}\cdot \textbf{Z} + \textbf{b} \cdot a_b(t)\ 
\text{where,}\  
    \textbf{Z} = \begin{bmatrix}\boldsymbol{\eta}\\ \dot{\boldsymbol{\eta}} \\ v_p \end{bmatrix}, \,\,\,
    \textbf{A} = \begin{bmatrix}0 & 1 & 0\\-\textbf{k}_o & -\textbf{c}_o & \boldsymbol{\theta}_o\\0 & -\frac{\boldsymbol{\Theta}^T\boldsymbol{\Phi}_o}{C_p}& -\frac{1}{C_p R_l}\end{bmatrix}, \,\,\,
    \textbf{b} = \begin{bmatrix}0 \\ \textbf{f}_o \\ 0 \end{bmatrix}
\end{equation}

Simulink \verb!ode45! solver is used to solve this equation, from which it is possible to estimate the generated voltage time history $v_p(t)$ employing an acceleration record as excitation. The built-in function \verb!ode45! is based on an explicit Runge-Kutta (RK) method, i.e. Dormand-Prince method \cite{DORMAND198019}. It combines the Runge-Kutta fourth and fifth order method and uses variable step size at each iteration to achieve the desired accuracy. Ultimately, the procedure allows to efficiently estimate (by considering only a small subset of vibration modes) the time history of the generated voltage for a given excitation. 

\subsection{Verification of the MOR Model for PEHs}
\label{S:2.3}

Although, the MOR method decreases the computational time required to estimate the output voltage, it has an approximation error compared to the full order model. Thus, it is necessary to seek a balance between accuracy and computational time. This issue is studied here and corresponds to the verification of the MOR model, i.e., how credible is the MOR model in comparison with the model with full degrees of freedom. A specific PEH is selected to analyse the impact of the number of vibration modes on processing time and accuracy. The geometrical parameters and the electromechanical properties of the device are given in tables \ref{noModes} and \ref{mat_prop_demarqui22}, respectively. The study is carried out adopting the excitation shown in Figure \ref{event02a} in Section 2, i.e. an event from the cable-stayed bridge acceleration database used in Sections \ref{S:Y} and \ref{S:5} as a case of study. The final outcome of the model is the amount of harvested energy, denoted by $E(\mathbf{x}\vert e)$, where vector $\mathbf{x}$ defines the PEH's geometry (i.e., $L$, $L_{pzt}$, $W$, $h_p$ and $h_s$) and $e$ is the event. This notation explicitly indicates that the harvested energy depends on the geometry of the device and the excitation time history (event). Ultimately, the amount of the produced energy is estimated from the numerical integration of the power signal $P(t, \mathbf{x}\vert e) = v^2(t, \mathbf{x}\vert e)/R_l$, i.e.
\begin{equation}
    \label{powersignal}
    E(\mathbf{x}\vert e) = \int_{t_1}^{t_2}P(t, \mathbf{x}\vert e) dt = \int_{t_1}^{t_2}\frac{v^2(t, \mathbf{x}\vert e)}{R_l} dt,
\end{equation}
where $v(t, \mathbf{x}\vert e)$ denotes voltage for device $\mathbf{x}$ and event $e$.

\begin{table}[h!]
\centering
\caption{Geometrical characteristics of the PEH employed in the verification of the MOR model.}
\label{noModes}
\begin{tabular}{@{}lrl}
\hline
Length of substructure layer ($L$)                             &200             & [mm]\\
Length of piezoelectric layer ($L_{pzt}$)                   &200             & [mm]\\
Width ($W$)                                              &200             & [mm]\\
Piezoelectric layer thickness ($h_p$)                         &0.25            & [mm]\\
Substructure layer thickness ($h_s$)                          &0.50            & [mm]\\
\hline
\end{tabular}
\end{table}

\begin{table}[h!]
\caption{Substructure and piezoelectric mechanical and electromechanical properties for the PEH employed in the verification of the MOR model.}
\label{mat_prop_demarqui22}
\centering
\begin{tabular}{@{}lrl}
\hline
Substructure Young's modulus, $E_s$  & 105 &[GPa]\\
Substructure density, $\rho_s$  & 9000 & [kg/m\textsuperscript{3}]\\
Substructure Poisson's ratio, $\nu$ & 0.3 \\ 
Piezoelectric layer density, $\rho_p$  & 7800 & [kg/m\textsuperscript{3}]\\
Permittivity  & 1800 $\times \epsilon_0 $ & [nF/m]\\
$c_{11}^E, c_{22}^E$  & 120.3 & [GPa]\\
$c_{12}^E$  & 75.2 & [GPa]\\
$c_{13}^E, c_{23}^E$  & 75.1 & [GPa]\\
$c_{33}^E$  & 110.9 & [GPa]\\
$c_{66}^E$  & 22.7 & [GPa]\\
$e_{31}^E, e_{32}^E$  & -5.2 & [C/m\textsuperscript{2}]\\
$e_{33}$  & 15.9 & [C/m\textsuperscript{2}]\\
\hline
\end{tabular}
\end{table}

\begin{figure}[h]
\centering\includegraphics[width=0.85\linewidth]{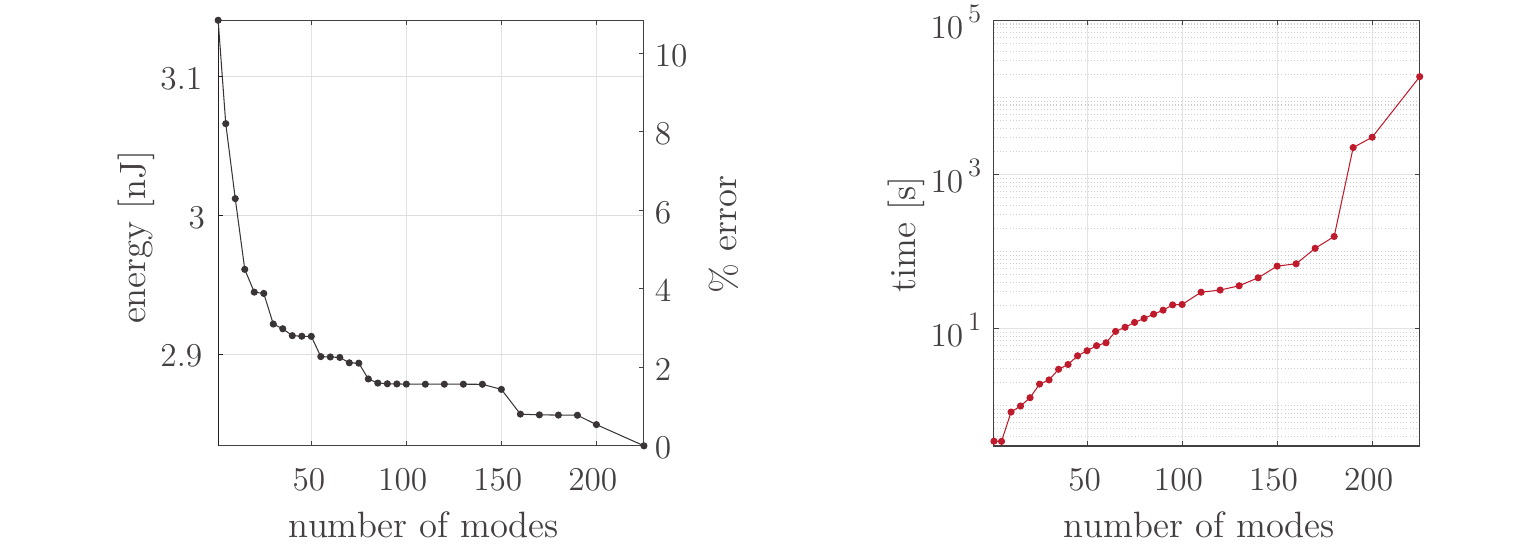}
\caption{Estimated energy harvested and relative simulation error (with respect to the full model) for MOR model with different number of modes (left). Computational time for MOR model with different number of modes (right).}\label{ModeStudy}
\end{figure}

\begin{figure}[h]
\centering\includegraphics[width=0.8\linewidth]{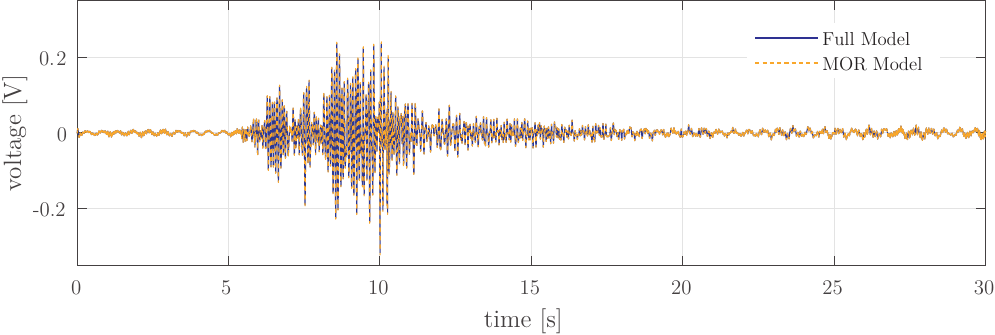}
\caption{Estimated voltage signal for the event 2, using the full model ($K = 225$) and the MOR model with 30 modes ($K=30$).  The computational time of MOR model is 3 sec, while the full model requires 1.9$\times$10$^4$ sec.}\label{event02v}
\end{figure}

Figure \ref{ModeStudy} shows the harvested energy and its relative error (with respect to the full model of 225 degrees of freedom) for different number of modes. On the other hand, in Figure \ref{ModeStudy}, the variation of computational time with respect to the number of modes is shown. As expected, reducing the number of modes decreases both, the computational time and accuracy. Therefore, thirty vibration modes will be considered in all numerical studies in this work since it provides the error of less than 3\% and a computational time of less than 2 seconds. The voltage signal estimated by the model with thirty modes for the event 2 (presented in Figure \ref{event02a}) is compared with the voltage signal estimated by the full model in Figure \ref{event02v}, where a good agreement between the two results can be seen. Note that the computing time of the MOR model is 3 sec, in contrast to the full model which takes 1.9$\times$10$^4$ sec. 

\subsection{Experimental Validation of the Model for PEHs}
\label{S:2.4}
In this section the MOR model is compared with experimental data obtained from analysing a commercial bimorph PEH. Hence, the validation answers the following question: how adequate are the MOR models to describe the physics involved in the problem. Note that the verification performed previously answers a different question: what is the accuracy of the MOR model with respect to the full model. In the experiment performed, the harvester is attached to an electromagnetic shaker such that any given excitation can be imposed. The harvester's piezoelectric layers are connected in series to a single electrical resistance of 1000 $\Omega$, forming an electrical circuit. An acquisition system is set to record the voltage $v(t)$ at the electrical resistant and the excitation acceleration $a_b(t)$ at the base of the PEH. This experimental setup is implemented following the directions given in \cite{peralta2019experimental}. 

Table \ref{T1} presents the electromechanical properties and the geometric characteristics of the PEH used in the validation. The identification of these properties for the commercial PEH was carried out using the Bayesian inference framework proposed by Peralta et al. in \cite{peralta2020bayesian}. 

\begin{table}[t]
\centering
\caption{Substructure and Piezoelectric mechanical and electromechanical properties of the PEH used in the experimental validation.}
\label{T1}
\begin{tabular}{@{}lrl}
\hline
Model Parameters& & \\
\hline
Substructure Density ($\rho_s$)         &7725.7            & [kg m$^{-3}$]\\
substructure Young Module ($Y_s$) 	                    &61.9	        & [GPa]\\
Compliance at constant electric field ($s_{11}^E$)      &15.2           & [pN$^{-1}$m$^{-2}$]\\
Piezoelectric Strain
Constant ($d_{31}$)                                     &292.8            & [pC N$^{-1}$]\\
Permittivity at Constant Stress ($\varepsilon_{33}^T/ \varepsilon_o$)  & 10299.1 & [F m$^{-1}$]\\
Piezoelectric Density($\rho_p$)                         &7567.9           & [kg m$^{-3}$]\\
Length ($L$)                                            &23.9             & [mm]\\
Width($W$)                                              &10.3             & [mm]\\
Piezoelectric Thickness ($h_p$)                         &0.245            & [mm]\\
Substructure Thickness ($h_s$)                          &0.232            & [mm]\\
Damping Ratio ($\zeta$)                                 &0.0126          &\\
\hline
$\varepsilon_o=8.854$x$10^{-12}$\\
\end{tabular}
\end{table}

\begin{figure}[h!]
\centering\includegraphics[width=0.7\linewidth]{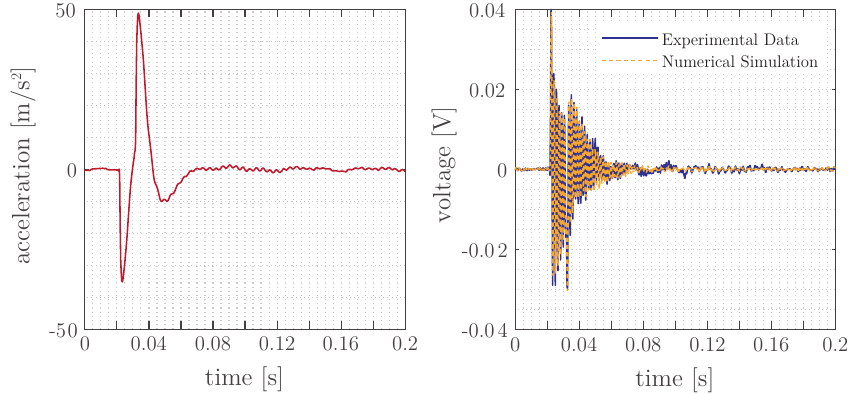}
\caption{Experimental validation of the numerical model of the PEH. Top figure presents the experimental acceleration at the base of the PEH. Bottom figure compares the numerical prediction and the experimental voltage signals.}\label{validationTI}
\end{figure}

Figure \ref{validationTI}a shows the pulse acceleration (excitation) recorded at the base where the commercial PEH is clamped. On the other hand, in Figure \ref{validationTI}b, the blue line corresponds to the experimental measurement of the voltage generated from the pulse excitation. Subsequently, the pulse acceleration is used as excitation in the numerical simulation (using the MOR model with $K=30$), and the respective response is plotted in Figure \ref{validationTI}b with a yellow dotted line. As can be seen in Figure \ref{validationTI}b, the numerical simulation agrees with the experimental results indicating the suitability of the proposed model to estimate the voltage response in the PEH. 

\section{Preliminary Studies}
\label{S:Y}

In this section, a preliminary analysis is carried out to explore the design space and compare optimal geometries, obtained from the parametric studies with those obtained from the event-based optimisation. 

\subsection{Parametric Studies}\label{opt_param_studies}

We assume that the geometry of a harvester is defined by five parameters, as shown in Figure \ref{PEH-scheme}: the total length $L$, the aspect ratio $R = W/L$ (relation between the width $W$ and the total length $L$), dimensionless piezoelectric length $l = L_{pzt}/L$ (relation between the length of the piezoelectric layers $L_{pzt}$ and the total length $L$), the total thickness $h = h_{s} + 2h_p$, and the dimensionless piezoelectric thickness $H = h_p/h$ (relation between the thickness of a piezoelectric layer $h_p$ and the total thickness $h$). A bronze substructure and two PZT 5A piezoceramic sheets connected in series are considered. The corresponding material properties are given in Table \ref{mat_prop_demarqui22}.

The studies conducted in this section are based on three design scenarios for parameters $(L, R, l, H)$, fixing two parameters and considering the other two as design variables. The reason for this is to help the visualisation and facilitate the interpretation of the results. In particular, the designs start from a basic design that considers a square shape ($R$=1),  a constant total thickness ($h = 1$mm), fully covered ($l$=1), and all layers with the same thickness ($H$=0.25). The three design cases are detailed next.
\begin{itemize}
    \item {\bf Design 1}: $H = 0.25$, $l = 1$, the design variables are: $L\in [10, 50]$ cm and $R\in [0.3, 1]$
    \item {\bf Design 2}: $R = 1$, $l = 1$, the design variables are: $L\in [10, 50]$ cm and $H\in [0.05, 0.45]$
    \item {\bf Design 3}: $R = 1$, $H = 0.25$, the design variables are: $L\in [10, 50]$ cm and $l\in [0.1, 1]$
\end{itemize}

In the first instance, a parametric study is carried out to understand the dynamic behaviour of harvesters with different geometries to justify the choice of search space and have an understanding of the potential best candidates before performing the full optimisation. 

In this context, an issue that requires special attention is the choice of the external electrical resistance ($R_l$) due to its important impact on power generation, as it has been discussed in other works \cite{hurtado2022shape,junior2009electromechanical,erturk2008distributed, paquin2010improving}. In that sense, the electrical resistance $R_l^*$ is chosen to maximise the value of the power FRF at the first resonance frequency $\omega_o$. Such optimisation problem can be expressed as
\begin{equation}
    \label{OptimalRL}
         R_l^*= \text{arg}\,\max\limits_{R_l \in \mathbf{R}_l} \, \vert H_p(\omega_o, R_l\vert \mathbf{x}) \vert 
\end{equation}
where $H_P(\cdot)$ is the voltage FRF defined in equation (\ref{HP_equation}), and the vector $\mathbf{x}$ defines the device geometry ($\mathbf{x}=[L\,\,R]$ for Design 1, $\mathbf{x}=[L\,\,H]$ for Design 2, and $\mathbf{x}=[L\,\,l]$ for Design 3). Note that the power FRF is evaluated at the first resonance frequency $\omega_o$. However, $\omega_o$  depends on the value of the external electrical resistance $R_l$, hence, $\omega_o$ is considered as a function of $R_l$. Therefore, the value of $H_p(\omega_o(\Bar{R_l}), \Bar{R_l}\vert \mathbf{x})$ is calculated numerically and the optimisation problem, equation (\ref{OptimalRL}), is solved using the Nelder-Mead Simplex Method \cite{doi:10.1137/S1052623496303470}. The procedure is then used in the parametric study and the subsequent optimisation processes in order to have consistency between the results.

\begin{figure}[h]
\centering\includegraphics[width=0.85\linewidth]{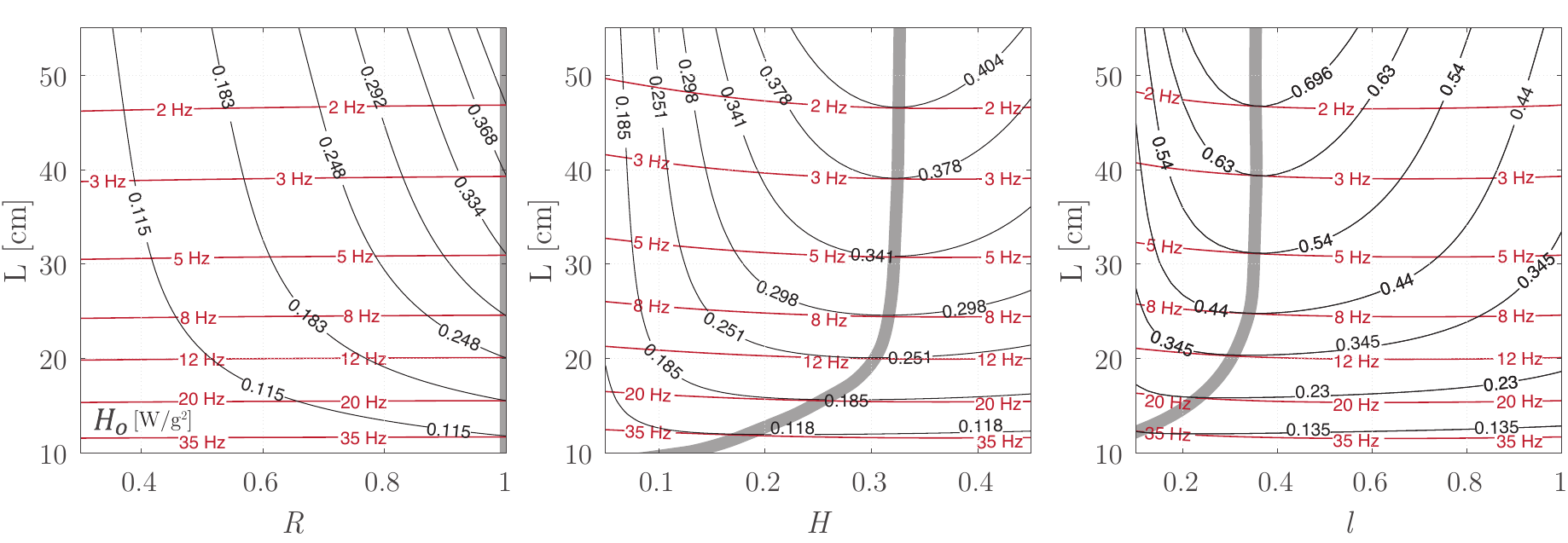}
\caption{Isocurves for peak amplitude $H_o$ of power FRF (black lines) and isocurves for natural frequencies $\omega_o$ (red lines) for the three design scenarios studied. The grey lines correspond to optimal geometries for any given natural frequency $\omega_o$.}\label{Isocurve-Fig}
\end{figure}

The results of parametric studies for the three design scenarios are shown in Figure \ref{Isocurve-Fig}, where the effect of the size, shape and distribution of the piezoelectric material on the natural frequency $\omega_o$ and its respective power FRF value $H_o$ are studied. The power FRF is defined in the equation (\ref{HP_equation}). The isocurves for $\omega_o$ and $H_o$ in the ($R$, $L$) plane for Design 1 are represented in Figure \ref{Isocurve-Fig}a with a red and black line, respectively. As expected, $\omega_o$ decreases with increasing length $L$ due to the reduction in stiffness and the increase in mass. On the other hand, $\omega_o$ is almost independent of $R$. Concerning the maximum power $H_o$, the increase of $R$ and $L$ favours the generation of energy due to the increase in the device area and the amount of mass that increases the inertial effect.

Figure \ref{Isocurve-Fig}b shows $\omega_o$ (red line) and $H_o$ (black line) isocurves in the ($H$, $L$) plane for fully covered square devices (Design 2). Note that when the dimensionless thickness $H$ is increased for any $L$, a slight increase in the natural frequency $\omega_o$ is observed. It must be taken into account that the piezoelectric material has a lower density and Young's modulus than the substructure. This implies that the mass and stiffness of the device decrease when $H$ increases. However, the change in stiffness is the one that dominates the dynamic behaviour resulting a decrement in $\omega_o$. Regarding the generated power $H_o$, it increases with $L$ mainly due to the increase in the PZT area. On the other hand, there is an increment and subsequently a reduction of $H_o$ with increasing $H$. This behaviour is the result of different factors. First, increasing the thickness of the piezoelectric layer decreases the inertial effect of the structure and consequently, there is less mechanical forcing over the PEH. Furthermore, the electromechanical coupling and the capacitance of the device are inversely proportional to the thickness $h_p$. In addition, a higher value of $H$ brings an increment in the flexibility of the device, which favours the increment in deformations during the excitation.

The isocurves of $\omega_o$ and $H_o$ in the ($l$, $L$) plane for Design 3 are shown in Figure \ref{Isocurve-Fig}c. As seen in the Figure, the natural frequency $\omega_o$ slightly increases when $l$ decreases because of the increase in inertia in the area of the device's tip. On the other hand, similar to the previous cases, the natural frequency is proportional to $L$ despite increasing damping ratios. Regarding $H_o$, it increases and then decreases as $l$ decreases. It is due to the increase in the excitation force related to the greater inertia of the device. However, the decrease in $l$ implies a decrease in the area of the piezoelectric layers, responsible for the subsequent decrease in power $H_o$.

This parametric study reveals the importance of properly selecting the geometry of the PEH in the chosen search space in order to maximise power output. Indeed, it is possible to identify geometries that maximise the generated power $H_o$ for different natural frequencies $\omega_o$ (this is equivalent to finding the intersection of the $\omega_o$-isoline with the $H_o$-isoline with the highest value of $H_o$). These geometries form the lines coloured in grey shown in Figure \ref{Isocurve-Fig}. Note that these results are independent of the source of vibration, e.g. the bridge. This means, the designer can take the dominant vibration frequency of a particular bridge and select the geometry of the PEH based on the grey lines shown in Figure \ref{Isocurve-Fig}. Note that in previous works, the design process remits to tuning the device with one of the resonance frequencies of the bridge, so these grey lines serve as a benchmark to compare event-based optimisation results with the design process used in previous work.

\subsection{Event-Based Optimisation}\label{opt_event_based}
In this section we analyse the energy harvested from events. The optimisation problem for a single event is formalised as
\begin{equation}
    \label{optimal}
         \mathbf{x^*}= \text{arg}\,\max\limits_{\mathbf{x} \in \mathbf{X}} \, \vert E(\mathbf{x}\vert e) \vert
\end{equation}

The objective function $E(\mathbf{x}\vert e)$ is the energy harvested from a specific event $e$ by a PEH whose geometry is defined by vector $\mathbf{x}$. Here, an event is defined by peaks of accelerations, such that accelerations greater than a certain threshold indicate the presence of an event in the continuous acceleration record of bridge vibrations. An event can be caused by environemental conditions such as wind excitation or by the passage of vehicles over the bridge. In this work, no distinctions are made regarding the cause.

Optimisation is implemented for the three design scenarios in order to validate the results and study the variability in the optimal geometries resulting from various events. The optimisation for each event is carried out based on the  Particle Swarm Optimisation (PSO) method \cite{poli2007particle,perez2007particle,kang2012damage} employing thirty particles. PSO is chosen due to its adequate performance for non-convex problems. PSO is an iterative method where a population of candidate solutions (or particles) moves in the search space towards the best solution. Each particle is associated with a position and a velocity, which are redefined in each iteration based on the individual and collective experience. PSO is expected to be an efficient optimisation algorithm being fast and computationally cheap; also it can be parallelised to further increase the computational efficiency. 

\begin{figure}[h!]
\centering\includegraphics[width=0.85\linewidth]{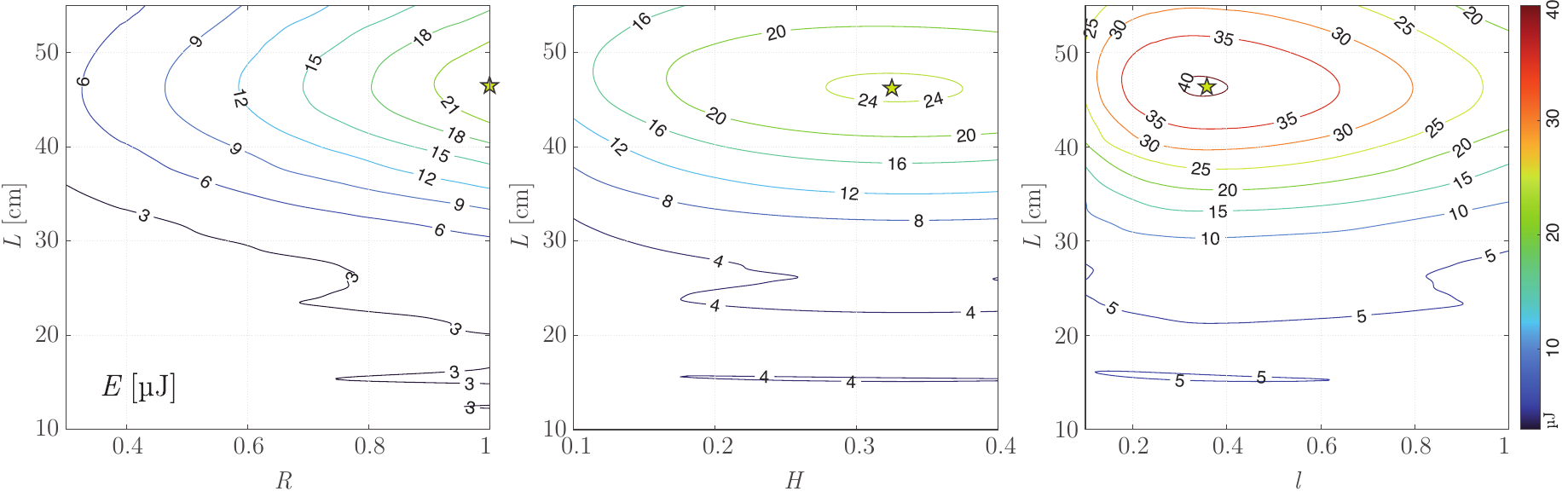}
\caption{Isocurves for harvested energy from event 2 and the optimal geometry obtained from the PSO method: Design 1 (left), Design 2 (middle), Design 3 (right).}\label{ISOiterations}
\end{figure}

\begin{figure}[h!]
\centering\includegraphics[width=0.85\linewidth]{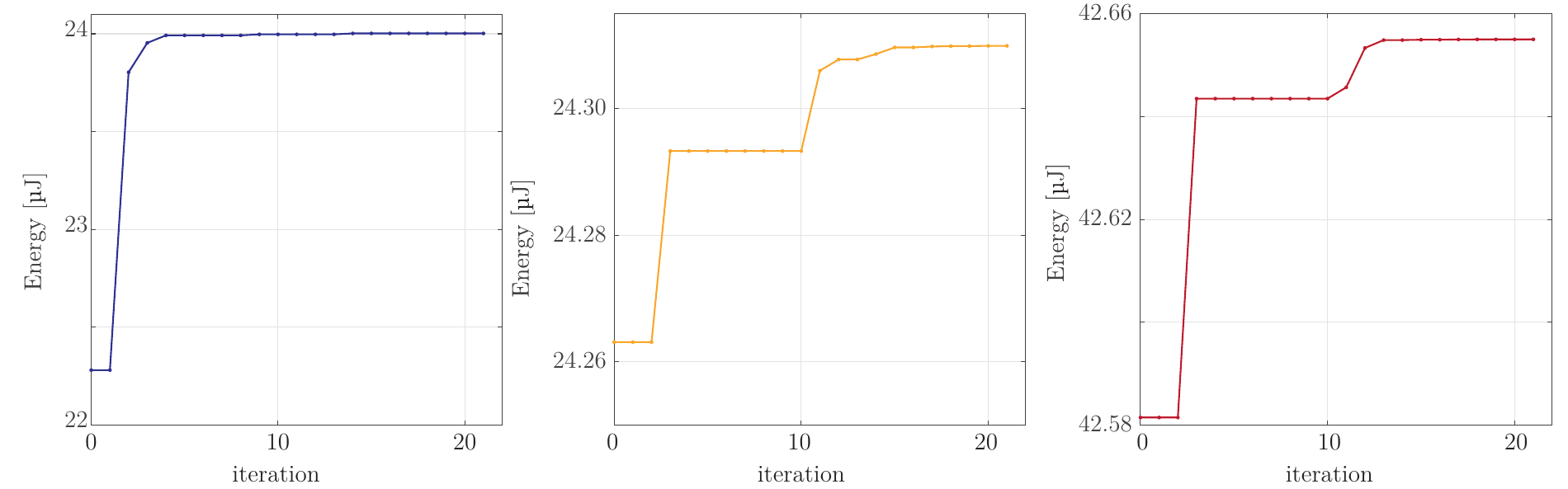}
\caption{Convergence process of the PSO algorithm: Design 1 (left), Design 2 (middle), Design 3 (right).}\label{PSOiterations}
\end{figure}

Figure \ref{ISOiterations} presents the iso-curves of the energy harvested from event 2 for the three design scenarios. In addition, the optimal geometries are marked with a star in the three figures. It is observed that the optimal geometry, obtained by the PSO, corresponds to the position estimated by the isocurves, which demonstrates that the method converges to a correct value (global maximum). Convergence of the objective function is shown in Figure \ref{PSOiterations}. The figure shows that twenty iterations are sufficient to find the optimal value.

\begin{figure}[h!]
\centering\includegraphics[width=0.85\linewidth]{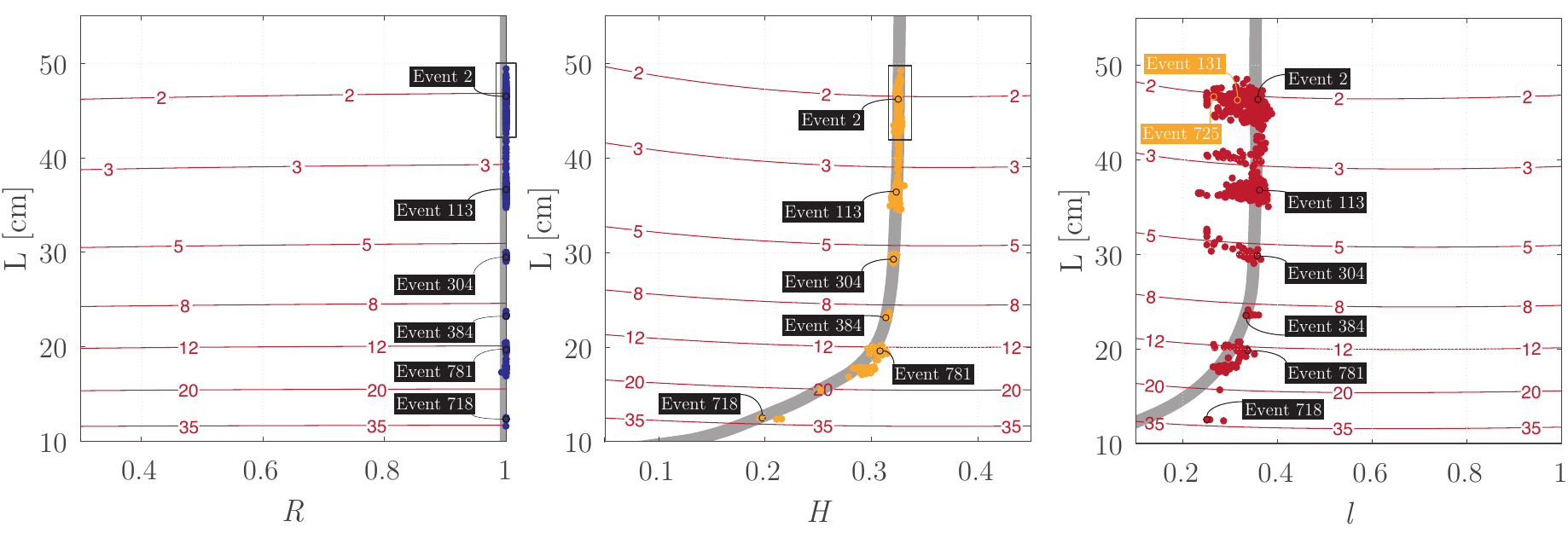}
\caption{Isocurves for fundamental frequencies $\omega_o$ (red lines) and optimal geometries obtained from peak amplitude $H_o$ of power FRF (grey lines). Optimal geometries for 1000 events for $(a)$ Design 1 (blue dots), $(b)$ Design 2 (yellow dots), $(c)$ Design 3 (red dots). The geometries related to events 2, 113, 304, 384, 781 and 718 are identified in three figures; also geometries for events 113 and 728 are denoted in the right figure.}\label{ShapeOpt}
\end{figure}

Figure \ref{ShapeOpt} shows the 1000 optimal PEH designs (one for each event) marked with blue, yellow and red dots for Designs 1, 2 and 3, respectively. Additionally, the figure shows the isocurve of the natural frequencies for different geometries (red line) and optimal designs from tuning the PEH's resonance frequency with some specific frequency (grey area) identified in the parametric studies. 

\begin{figure}[h!]
\centering\includegraphics[width=0.90\linewidth]{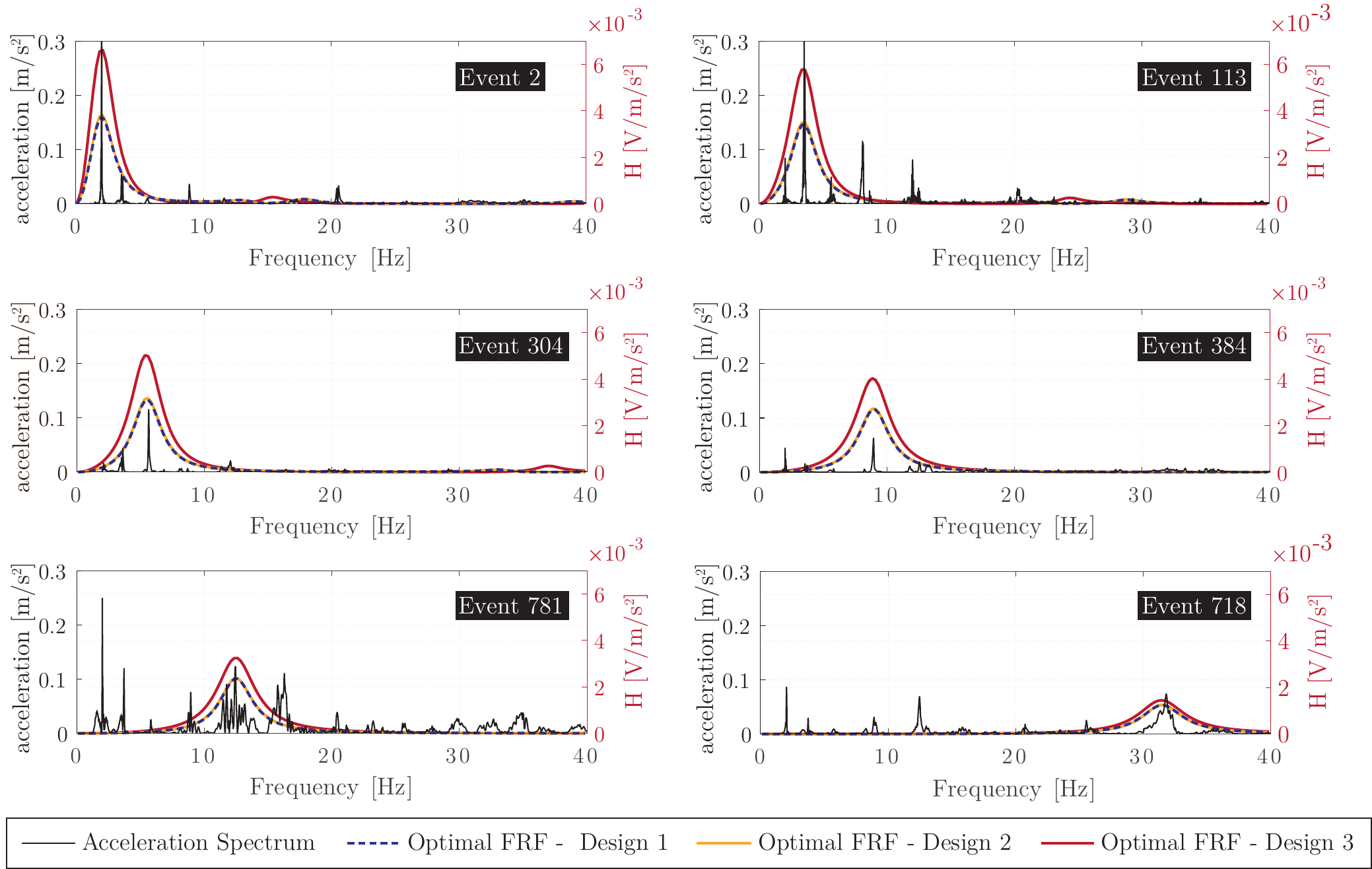}
\caption{Comparison between the Fourier spectrum of the event acceleration signal and the FRFs of the optimal geometries obtained from the three design scenarios for events 2, 113, 304, 384, 781 and 718.}\label{FRFevents}
\end{figure}

The results in Figure \ref{ShapeOpt} present high variability with respect to the natural frequency. This is explained by the differences in the spectral characteristics of the events. Figure \ref{FRFevents} presents the comparison of FRFs of six optimal designs and the Fourier acceleration spectrum of their respective event for the three design scenarios. The six optimal designs are identified in Figure \ref{ShapeOpt} with the purpose to be compared, which correspond to events 2, 113, 304, 384, 781 and 718. First, note that the fundamental frequency of the piezoelectric devices is tuned with the predominant frequency of the acceleration spectrum. However, the acceleration peak is not always the one with the highest amplitude, such as the ones in events 781 and 718 where the devices' fundamental frequency is tuned with the widest acceleration spectrum peak. Also, it is important to mention that there is consistency between the optimal FRFs for the three design scenarios since the fundamental frequencies coincide in the six studied events.

\begin{figure}[h!]
\centering\includegraphics[width=0.9\linewidth]{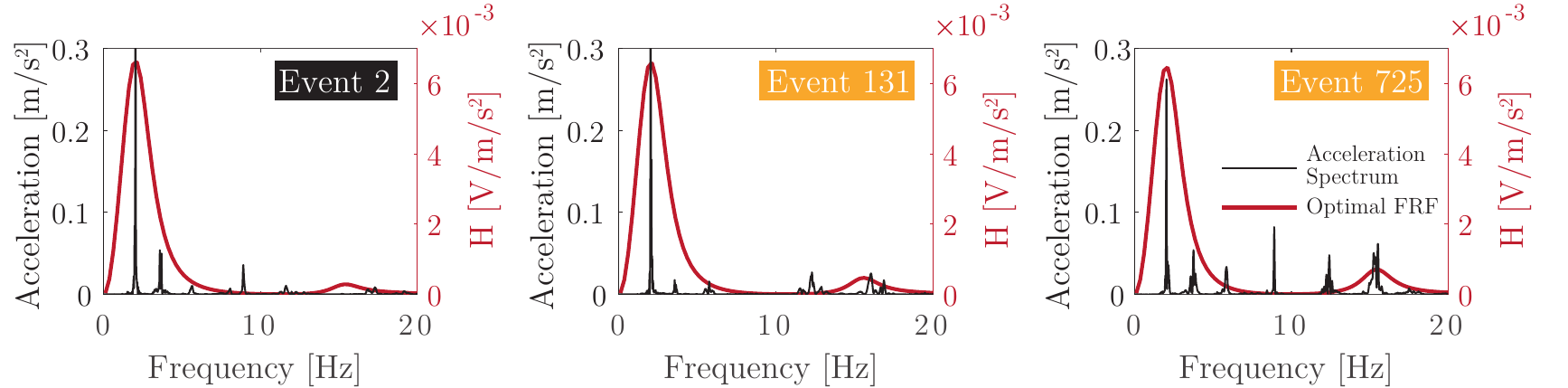}
\caption{Comparison between the Fourier spectrum of event acceleration signal and the FRFs of the optimal geometries (Design 3) from the three shape optimisation cases for events 2, 131 and 725.}\label{FRFevents2}
\end{figure}

Note that optimal designs within the grey line in Figure  \ref{ShapeOpt} are those that generate the higher energy from their first vibration mode. This is valid for all PEHs in Designs 1 and 2 while they present a higher variability in case 3. To understand the higher dispersion of the optimal geometries for Design 3, Figure \ref{FRFevents2} presents the comparison between the FRFs of three optimal designs and the Fourier acceleration spectrum of their respective event. The three optimal designs are identified in Figure \ref{ShapeOpt}-c, which correspond to events 2, 131 and 725. The three devices have the first natural frequency of 2 Hz. The high dispersion is explained by analysing the second natural frequency of the FRF. The main difference between Designs 1 and 2, against Design 3 is in the second natural frequency, which is more sensitive to change in parameter $l$ rather than parameters $R$ or $H$. The change in parameter $l$ allows increasing the energy generation from the second mode of vibration, slightly decreasing the energy generation from the first one, leading to a higher energy overall. Therefore, the main conclusion from this study is that according to the spectral characteristics of an event, in some cases it is better to use devices that harvest energy only around the first natural frequency, while in others it is more efficient to harvest energy simultaneously from the first two natural frequencies.

The results of this section can be summarised as follows. The studies reveal the importance of PEH geometry on energy output. In this sense, the PSO method demonstrated its capability to converge to the global maximum correctly and efficiently. In addition, high variability was observed in the optimal geometries when using a large number of events, which is explained by different spectral characteristics of events. From these findings we can conclude that the optimisation of devices considering a large number of events gives it greater robustness than the design obtained by simply tuning the natural frequency of the device with a target value.

In practice, we are interested in designing one or a limited number of piezoelectric devices for a continuous energy generation; thus, it is not feasible to design a device for each event or perform an optimisation process in long time windows due to computational limitations. To overcome this limitation, in the next section we introduce the framework for designing piezoelectric devices for a continuous energy generation.

\section{Optimisation of PEHs for Bridge Applications}
\label{S:3}

\begin{figure}[h]
\centering\includegraphics[width=0.8\linewidth]{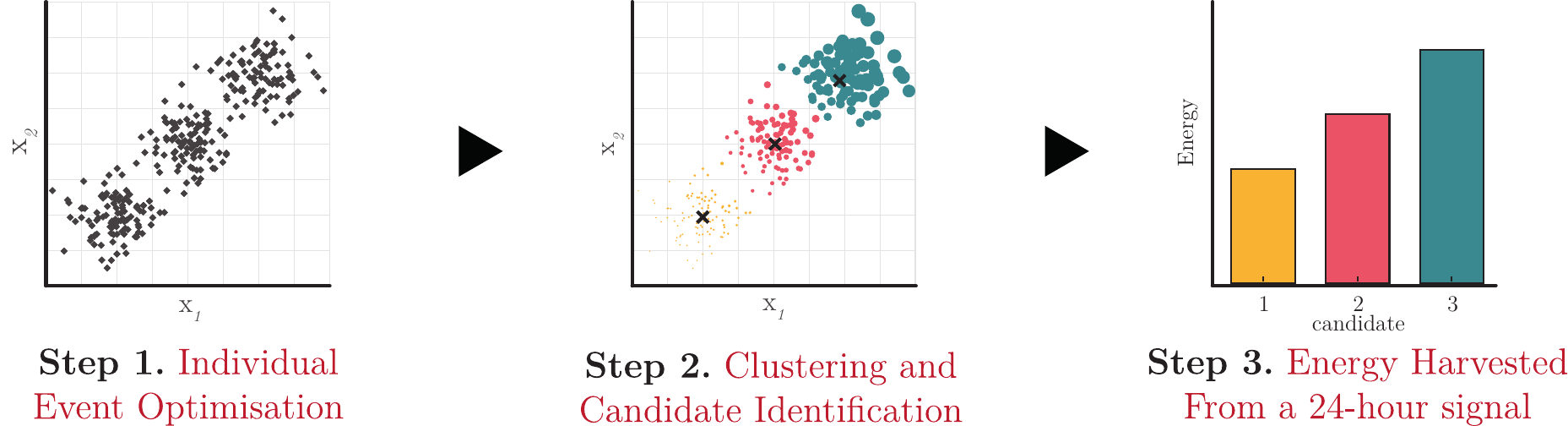}
\caption{Overview of the proposed optimisation framework.}
\label{steps}
\end{figure}

An optimisation framework for designing a PEH to maximise the power generation for an in-service bridge requires to record the acceleration of the bridge at location of the PEH. This data should be recorded in a time window large enough that can be taken as a representative behaviour of the bridge response. The recorded acceleration is taken as the base excitation of the PEH. The vibration response of the bridge is recorded continuously, and subsequently, events are extracted using a threshold criterion. After the identification of events, it is possible to apply the proposed optimisation procedure. The procedure consists of three parts: ({\bf Step 1}) the optimisation of the PEH for each event, ({\bf Step 2}) the clustering of optimal geometries, and ({\bf Step 3}) the estimation of the energy harvested for long time windows. The scheme of the procedure is shown in Figure \ref{steps}, and the steps are detailed next.

The aim of {\bf Step 1} is to infer the variability in optimal geometries considering a relatively large number of events. Shape optimisation is independently performed for each event (we call it \textit{Individual Event Optimisation}) as it was done in Section \ref{opt_event_based}. A total of $Ne$ events are considered and for each event maximum of $E(\mathbf{x}\vert e)$ (as defined by equation (\ref{optimal})) is obtained. In the schematic view shown in Figure \ref{steps} (Step 1), each dot represents the optimum design in terms of variables $\{$x$_1$, x$_2\}$, independently obtained for each event. The objective function could be modified, for example to design PEHs considering the energy per unit of area.

The objective of {\bf Step 2} is to deal with the variability of the optimal geometries and to derive optimal candidates for continuous power generation. This step is called \textit{Clustering and Candidate Identification}. The optimal geometries from each event tend to have a natural frequency close to one of the bridge's resonance frequencies. Thus, it is possible to group them in clusters based on the geometrical parameters (design variables). In Figure \ref{steps} (Step 2), optimal designs are schematically grouped into three clusters, denoted by yellow, red and green colours with their centroids denoted by black crosses. In this work, $k$-means \cite{kanungo2002efficient, diez2016clustering} is selected to cluster the geometries due to its ease of implementation, efficiency, empirical success, and adaptability in clusters of different shapes and sizes. $k$-means is an unsupervised machine learning algorithm that assigns each $point$ (optimal design) in one $k$ group based on its characteristics (geometric parameters and harvested energy from event data). The clusters are represented by a centroid which features are adjusted iteratively, minimising the distance with the assigned elements in the group. The number of clusters $k$ needs to be set before starting the algorithm, however, an arbitrary selection does not guarantee the best clustering result. To deal with this issue, the Silhouette Method \cite{yuan2019research} is also included in this framework to find the optimal $k$ value. The Silhouette method computes coefficients that measure how much a $point$ is similar to its cluster compared to others, providing a metric to evaluate how well each object has been classified. Finally, once the optimal designs are grouped, the centroids are defined as candidates to be the optimal PEHs for continuous energy generation.

The last step, {\bf Step 3}, consists of assessing the \textit{Energy Harvested from a 24-hour acceleration signal} (or any other time window) for each candidate obtained in Step 2. The purpose of this step is to select the best option (among the candidates) for continuous energy generation. In the real life scenario, this is the preferred design because the device should generate the optimal amount of energy independently on the traffic on the bridge. In Figure \ref{steps} (Step 3), energy generated by three candidates (centroids identified in Figure \ref{steps} (Step 2)) is shown schematically.      



\section{Case Study: Optimisation of PEHs for an in-service Bridge}
\label{S:5}

The optimisation framework proposed in Section \ref{S:3} is now applied to the same cable-stayed bridge presented in the Section \ref{S:B}. The framework is expected to offer candidates to maximise the energy harvested in continuous operation. 1000 Events are considered in this case as a representative number for this bridge because it is the approximate number of events that occur during a week, even though the selected events are drawn from different periods. The PEH materials are chosen the same as in Section \ref{S:2.3}. A serial connection between the piezoelectric layers is considered, and the selected electrical resistance for each geometry is the one that maximises the peak of the FRF at the fundamental frequency. Five design parameters describe the geometry: total length $L\in$[10, 50] cm, aspect ratio $R\in$[0.3, 1], dimensionless piezoelectric length $l\in$[0.1, 1], dimensionless piezoelectric thickness $H\in$[0.05, 0.45], and total thickness $h$ which is assumed constant and equal to 1 mm. Design variables are defined as
\begin{equation}
    \mathbf{x} = \{R, L, l, H\}
\end{equation}

\begin{figure}[h!]
\centering\includegraphics[width=0.85\linewidth]{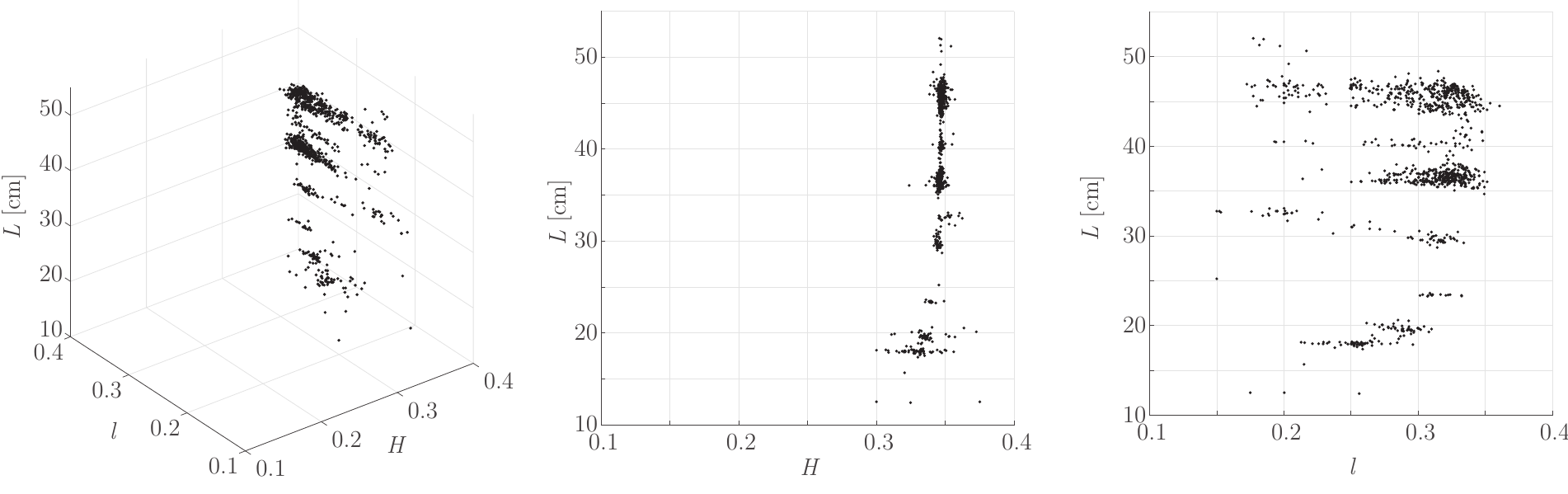}
\caption{Optimal geometries for 1000 events in the $L-l-H$ space (left), $L-H$ plane (middle) and $L-l$ plane (right).}
\label{OptimalGeometriesCS1}
\end{figure}

First, the individual event optimisation was performed for 1000 events. As a result, 1000 optimal geometries were obtained. It was observed that all optimal geometries correspond to $R = 1$, which is in agreement with the results in Section \ref{S:Y}, hence variable $R$ is omitted from the further analysis. Optimal designs are shown in Figure \ref{OptimalGeometriesCS1} in $L-l-H$ space and their projections in the $L-l$ and $L-H$ planes. From Figure \ref{OptimalGeometriesCS1}b it can be seen that the results present a high variability for the variable $L$ and low for the variable $H$. The high variance in $L$ is explained by its strong dependence on the natural frequency of the device, the high variability in the predominant frequencies in the events, and the fact that the devices tend to synchronise to the event's predominant frequencies. In Figure \ref{OptimalGeometriesCS1}c, it is possible to observe a variability in the variable $l$. This is explained by the fact that the devices tune their first two resonance frequencies with bridge's natural frequencies in some events. In particular, $l$ allows increasing the energy generation from the second mode of vibration, slightly decreasing the energy generation from the first one, leading to an overall higher harvested energy according to the specific spectral characteristic of the event.

\begin{figure}[h]
\centering\includegraphics[width=0.45\linewidth]{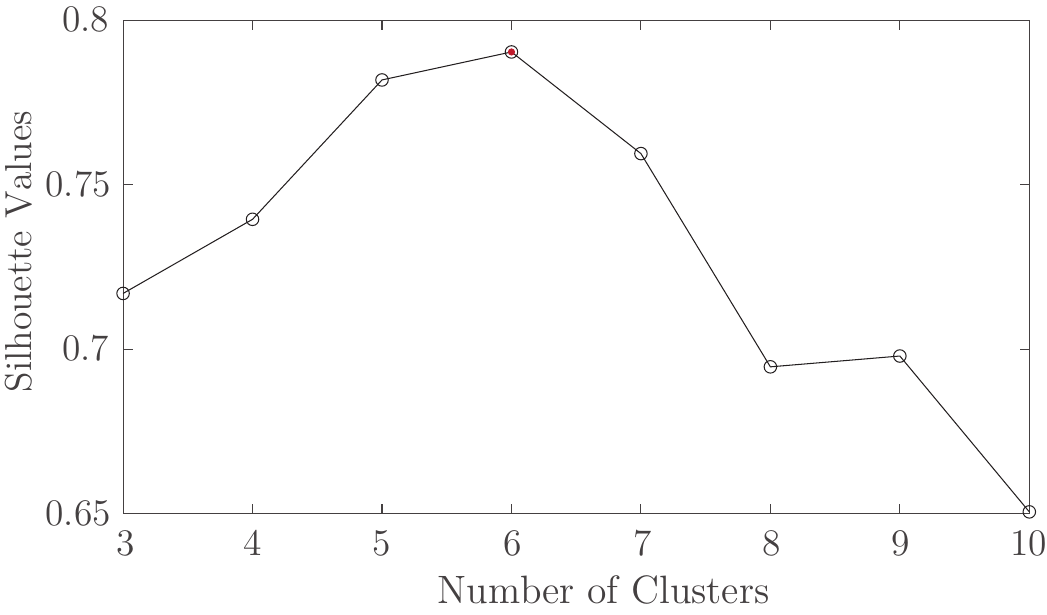}
\caption{Silhouette coefficient for different $k$-values applied to the 1000 optimal geometries.}
\label{SilhouetteCS1}
\end{figure}

\begin{figure}[h]
\centering\includegraphics[width=0.85\linewidth]{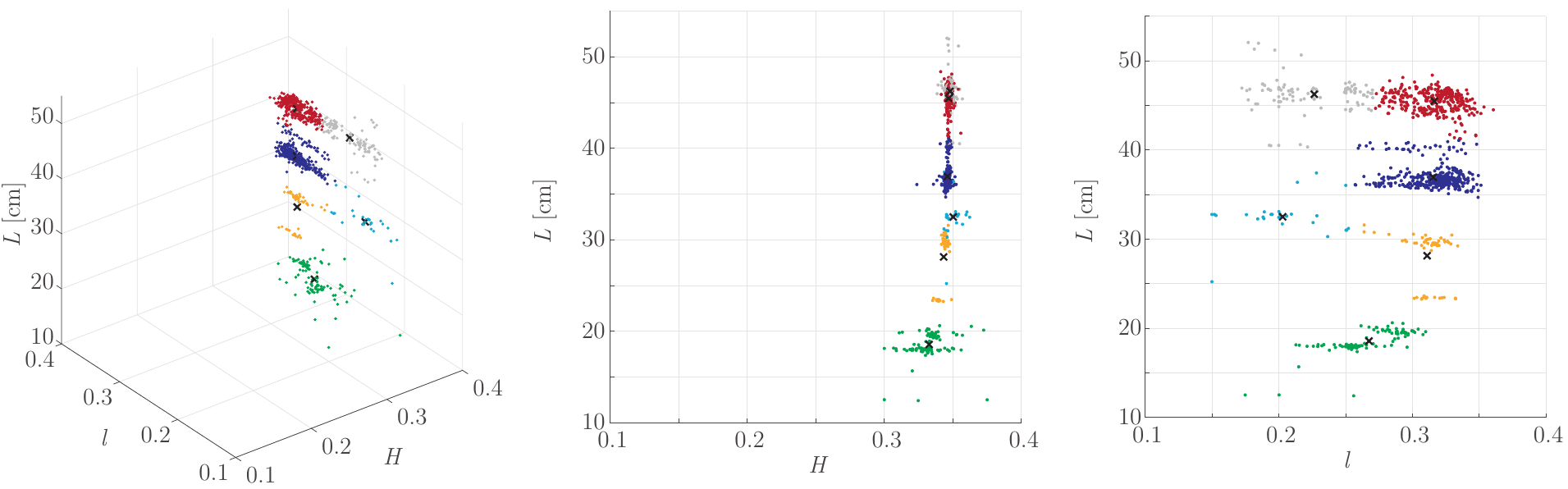}
\caption{Six clusters of the optimal geometries for 1000 events in the $L−l−H$ space (left), $L−H$ plane (middle) and $L−l$ plane (right). Black crosses denote candidates (centroid of each cluster).}
\label{clusterCS1}
\end{figure}

After the individual event optimisation, cluster identification is carried out where the optimal geometries are grouped based on the similarity between them, taking into account the geometric variables $L$, $l$ and $H$. The cluster number is determined based on the Silhouette coefficient, which allows evaluating how well each geometry has been classified in their cluster with respect to the others. The results for different cluster numbers are presented in Figure \ref{SilhouetteCS1}, where it is concluded that the optimal number is 6 clusters. Next, cluster identification is performed based on the $k$-means algorithm. The six clusters and their respective centroids are presented in Figure \ref{clusterCS1}. In this framework, the centroids are considered to be representative candidates of each cluster. Finally, the 1000 optimal geometries have been reduced to six candidates that will be evaluated for the continuous power generation.

\begin{figure}[h]
\centering\includegraphics[width=0.85\linewidth]{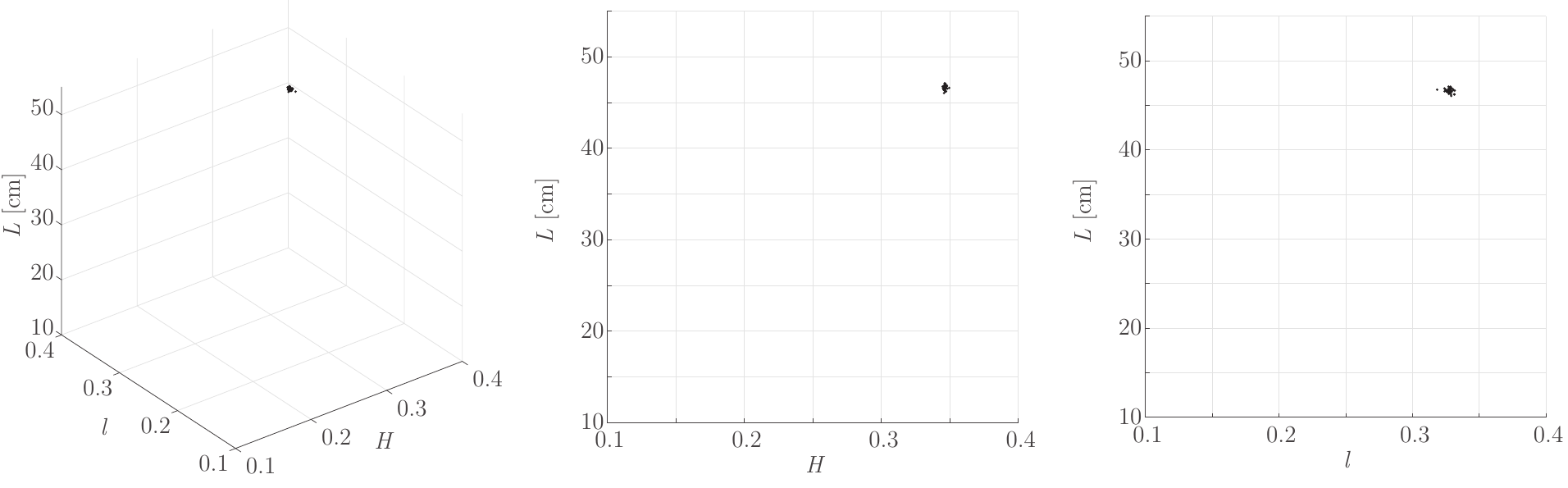}
\caption{Optimal geometries for 100 time windows of 30 seconds without events in the $L−l−H$ space (left), $L−H$ plane (middle) and $L−l$ plane (right).}
\label{noEventCS1}
\end{figure}

The cable stayed bridge studied in this work is generally not subjected to high traffic, so the frequency of occurrence of events is low. Therefore, the optimisation framework is also evaluated in time windows without events (i.e., small oscillations in response to environmental conditions) since this condition prevails over time. In addition, this analysis will function as a control group for the present case study. 100 time windows of 30 seconds without events are extracted from the same period considered in the 1000 event extraction. Then for each time window, the corresponding acceleration signal is used as an input in the optimisation problem. The results of such individual optimisation are shown in Figure \ref{noEventCS1}. Note that the optimal geometries present a lower variability, as expected, since the oscillations in the absence of traffic are consistent in time. Therefore, it seems reasonable to consider only one design candidate to evaluate it in the continuous operation and compare it with the other six candidates defined above.

\begin{figure}[h!]
\centering\includegraphics[width=0.95\linewidth]{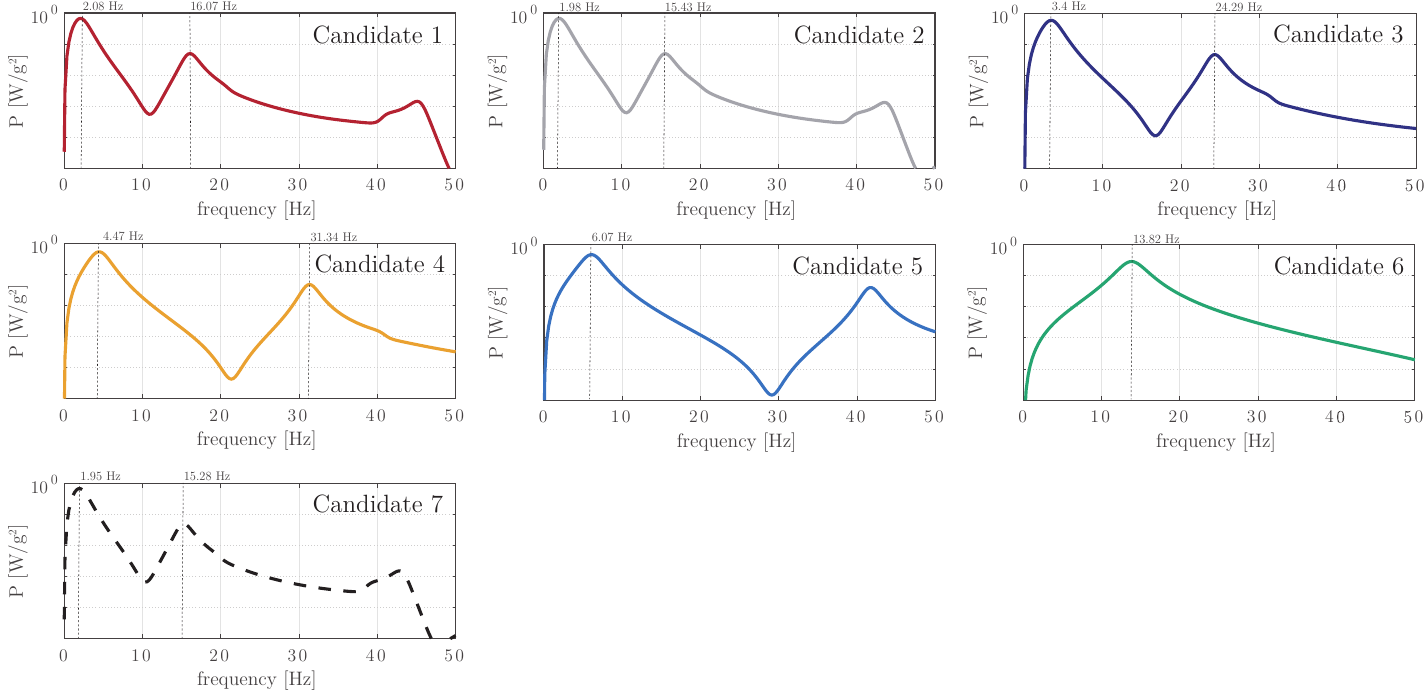}
\caption{Power FRF of the seven candidates to assess energy harvested from a continuous 24-hour signal.}
\label{FRFcanditates}
\end{figure}

\begin{figure}[h]
\centering\includegraphics[width=0.5\linewidth]{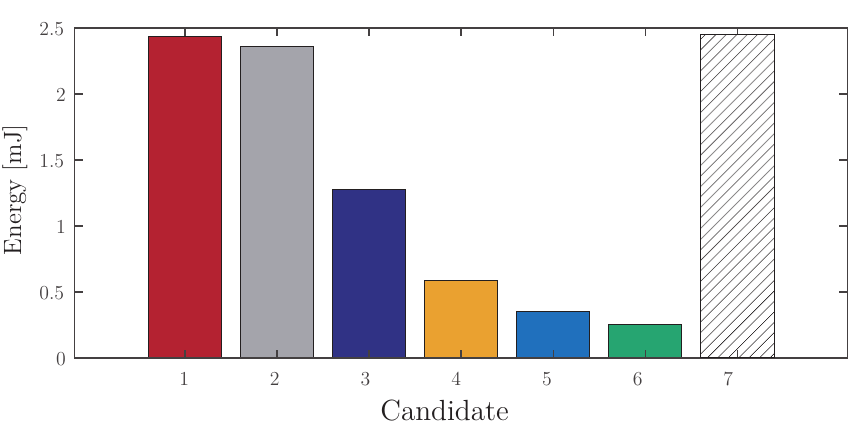}
\caption{Energy harvested from a 24 hour signal by the seven candidates.}
\label{energy24hrs}
\end{figure}

Finally, the seven PEH candidates are evaluated under 24 hour continuous operational condition. Figure \ref{FRFcanditates} presents their power FRFs, where it is possible to identify their natural frequencies. Note that these values coincide with the frequencies, where the optimal geometries from the optimisation in Section \ref{S:Y} are grouped. It is also important to note that the first two natural frequencies of candidates 1 and 2 coincide and only vary in their amplitude in the FRF, which agrees with the results observed and discussed previously in Design 3 of Section \ref{S:Y}. Next, Figure \ref{energy24hrs} presents the energy harvested from a 24 hour signal by the seven candidates. We can conclude that the candidates with the highest performance are the candidates whose first natural frequency is around 2 Hz, with candidate 7 being the one that generates the most energy. Note that candidate 7 is representative of event-less window optimisations. However, the energy harvested by the best candidate is only 0.47\% and 3.79\% higher than candidates 1 and 2, respectively. 

Although, the optimisation framework suggests that the optimal device is the one obtained from the optimisation of the windows without events, this cannot be generalised to all bridges since it depends on different factors such as the structure of the bridge, the vehicular characteristics and the frequency of the traffic, all of them impacting the events' occurrence rate. These factors are intrinsically included in the framework and are the ones that define dissimilar results for each context. In order to highlight the impact of the occurrence rate, it is decided to include Table \ref{CandidatevsCluster}. This table contains the expected energy harvested (in a time frame of 30s) by each candidate when they are excited with events of different clusters. For example, if candidate 1 is excited with an event belonging to cluster 2, then it is expected to produce 6.10$\mu$J in 30s. As it is expected, the higher energy for an event is produced by its correlative candidate, e.g., for an event of cluster 2, the best candidate is also the number 2. Also, it is important to note that the highest energy is harvested when candidate 6 is excited with an event belonging to cluster 6. However, candidate 6 in Figure \ref{energy24hrs} is the candidate that generates the lowest amount of energy accumulated in 24 hours. Contrary, it is possible to observe that the lowest amount of energy generated by any candidate occurs when no events are presented in the excitation (column 7 in Table \ref{CandidatevsCluster}). However, candidate 7 is the one that generates the most amount of accumulated energy in 24hrs. This situation is explained by the occurrence rate of an event belonging to a specific cluster. The last row of the table incorporates the mentioned occurrence rate. Here, it is observed that events of cluster 6 have an occurrence rate close to 0.5\%, while the occurrence rate for an event-less excitation is close to 95\%. As it is seen, the occurrence rate plays a determinant role in the optimisation. Furthermore, the information in this table could be used to estimate the energy produced by any candidate in any time frame. For example taking candidate 1, if the first row (corresponding to candidate 1) is weighted by the occurrence rate (last row) and the result is summed, then the expected energy harvested by candidate 1 in 30 sec is obtained. Then, the result could be linearly scaled to estimate the expected energy harvested in a different time frame.

\begin{table}[h!]
\centering
\caption{Expected energy harvested (in a time frame of 30s) by all candidates when they are excited with events belonging to different clusters. The energy is in $\mu$J. The last row corresponds to the occurrence rate of an event belonging to a particular cluster for the case study}
\label{CandidatevsCluster}
\begin{tabular}{cllllllll}
\cline{3-9}
\multicolumn{1}{l}{}                            &                        & \multicolumn{7}{c}{Event's Cluster}                                                                                                                                           \\ \cline{3-9} 
\multicolumn{1}{l}{}                            &                        & \multicolumn{1}{c}{1} & \multicolumn{1}{c}{2} & \multicolumn{1}{c}{3} & \multicolumn{1}{c}{4} & \multicolumn{1}{c}{5} & \multicolumn{1}{c}{6} & \multicolumn{1}{c}{7} \\ \hline
\multicolumn{1}{c|}{\multirow{7}{*}{\rotatebox{90}{Candidate}	}} & \multicolumn{1}{l|}{1} & 5.50                   & 6.10                   & 2.13                   & 1.40                   & 1.06                   & 5.43                   & 0.54                   \\
\multicolumn{1}{c|}{}                           & \multicolumn{1}{l|}{2} & 5.31                   & 6.26                   & 2.00                   & 1.37                   & 1.15                   & 6.28                   & 0.52                   \\
\multicolumn{1}{c|}{}                           & \multicolumn{1}{l|}{3} & 3.30                   & 3.84                   & 2.92                  & 1.68                   & 1.30                    & 3.56                  & 0.22                   \\
\multicolumn{1}{c|}{}                           & \multicolumn{1}{l|}{4} & 1.36                   & 1.61                   & 1.54                   & 1.89                   & 0.88                   & 1.87                   & 0.03                   \\
\multicolumn{1}{c|}{}                           & \multicolumn{1}{l|}{5} & 1.86                   & 2.29                   & 2.11                   & 1.74                   & 1.43                   & 2.56                   & 0.07                   \\
\multicolumn{1}{c|}{}                           & \multicolumn{1}{l|}{6} & 0.91                   & 3.66                   & 0.60                   & 0.76                   & 0.98                   & 7.76                   & 0.00                   \\
\multicolumn{1}{c|}{}                           & \multicolumn{1}{l|}{7} & 5.51                   & 6.05                   & 2.00                   & 1.34                   & 1.03                   & 5.42                   & 0.55                  \\ \hline
                  Occurrence rate$^*$        &  & 0.017                  & 0.006                  & 0.018                  & 0.003                  & 0.001                  & 0.005                  & 0.950\\
                           \hline
\multicolumn{9}{c}{$^*$Time of occurrence in a period of one week.}\\
\end{tabular}
\end{table}

Table \ref{harvestedEnergyComparation} compares the energy generated in this work with some previous studies. Although, the results coincide with the order of magnitude, it does not exceed the harvested energy from other studies. This is mainly due to the characteristics and ways the bridge-vehicle system is modelled in these works. In particular, the fact that in \cite{zhang2018experimental} the research was carried out in an experimental setup with a scaled bridge-vehicle system, allows the generation of energy to occur in higher frequencies, which benefits the efficiency of the devices. On the other hand, the data used in \cite{romero2021energy} corresponds to a railway bridge, where higher vibration levels are observed in comparison to vehicular bridges, due to the greater mass and speed of trains. The same happens in \cite{song2019finite}, where the numerical simulation is modelled on a bridge-train system.

\begin{table}[h!]
\centering
\caption{Comparison of the harvested energy from the present work to some previous studies.}
\label{harvestedEnergyComparation}
\begin{tabular}{lrll}
\hline
This work     & 7.8  & $\mu$J & for each passing vehicle \\
              & 2.5          & mJ     & in 24 hours              \\
Romero et. al \cite{romero2021energy} & 3.6          & mJ     & for each passing train   \\
              & 0.84 & mJ     & in 3.5 hours             \\
Song \cite{song2019finite}         & 462  & mJ     & for each passing train \\
Zhang et. al \cite{zhang2018experimental}   & 60-565       & $\mu$J & for each passing vehicle \\ \hline
\end{tabular}
\end{table}

Another aspect that was not considered in this work is the dependence of the optimisation results on the device's position on the bridge. In general, the device's position is associated with the different vibration modes of the bridge, so the amplitude associated with specific vibration frequencies is affected. In particular, the position selected in this case study favours the first mode of vibration and explains why a more significant number of optimal devices from the individual optimisation will cluster around 2 Hz. However, it is not necessarily the optimal position on the bridge and other positions should be studied.

\section{Conclusions}
\label{S:6}
This work presented a comprehensive study for the design of Piezoelectric Energy Harvesters for a bridge. The numerical PEH model was based on the Kirchhoff-Love plate theory and the generalised Hamilton’s principle for electroelastic bodies. The model was discretised by the isogeometric analysis and coupled with the Particle Swarm Optimisation algorithm. In order to reduce the computation burden, a Modal Order Reduction was performed. To estimate the voltage signal from an arbitrary input acceleration signal, efficient time integration based on Runge-Kutta method was used. Additional optimisation algorithm was applied to determine the optimal value of the electric resistance. The approach was validated against experimental measurements.
 
Preliminary investigations were carried out to understand the impact of considering different PEH designs on a real bridge. From the studies, it was deduced that tuning the fundamental frequency of the device with one resonance frequency of the bridge did not result in the optimal configuration if it was evaluated for different traffic events on the bridge. It revealed the importance of developing an alternative methodology to perform design optimisation. For that purpose, an optimisation framework based on the bridge events was proposed to design a PEH with maximum continuous power generation. The framework was applied to a real cable-stayed bridge. 1000 traffic passing events and 100 time windows without events (of 30 seconds) were considered as a representative database of the bridge. The optimal configuration was achieved from the non-event windows due to the study case's specific bridge and traffic characteristics.


\section*{Acknowledgements}
This research is supported by the Comision Nacional de Investigacion Cientifica y Tecnologica de Chile through the project CONICYT/FONDECYT/11180812. The authors also wish to thank CSIRO's Digital Productivity business unit, Data61 for providing the research data. The instrumentation and the field tests of this bridge have been planned and conducted by researchers at Data61 in collaboration with academics at University of New South Wales (UNSW) and Western Sydney University (WSU).

\bibliographystyle{model1-num-names}
\bibliography{reference.bib}

\end{document}